\documentclass[useAMS,usenatbib]{mn2e}

\newcommand{\be}{\begin{equation}}
\newcommand{\ee}{\end{equation}}
\newcommand{\bea}{\begin{eqnarray}}
\newcommand{\eea}{\end{eqnarray}}
\newcommand{\bfig}{\begin{figure}}
\newcommand{\efig}{\end{figure}}
\newcommand{\bfigs}{\begin{figure*}}
\newcommand{\efigs}{\end{figure*}}
\newcommand{\bt}{\begin{table}}
\newcommand{\et}{\end{table}}
\renewcommand{\vec}[1]{ {\bf #1} }

\usepackage[dvips]{graphicx}

\title[Simulations of galaxy formation with radiative transfer]
{Simulations of galaxy formation with radiative transfer:
Hydrogen reionisation and radiative feedback}
\author[M. Petkova and V. Springel]{Margarita Petkova$^1$\thanks{E-mail: mpetkova@mpa-garching.mpg.de} and
Volker Springel$^{2,3,1}$\thanks{E-mail: volker.springel@h-its.org}\\
$^1$Max-Planck-Institut f\"ur Astrophysik, Karl-Schwarzschild-Strasse 1, 
85748 Garching, Germany\\
$^2$Heidelberg Institute for Theoretical Studies,
  Schloss-Wolfsbrunnenweg 35, 69118 Heidelberg, Germany\\
$^3$Centre for Astronomy, Heidelberg University, M\"{o}nchhofstr. 12-14, 69120 Heidelberg, Germany}

\begin{document}

\date{Accepted 2010 October 29; Received 2010 October 18}

\pagerange{\pageref{firstpage}--\pageref{lastpage}} \pubyear{2010}

\maketitle

\label{firstpage}


\begin{abstract}
We carry out hydrodynamical simulations of galaxy formation that
simultaneously follow radiative transfer of hydrogen-ionising photons,
based on the optically-thin variable Eddington tensor approximation as
implemented in the {\small GADGET} code. We consider only star-forming
galaxies as sources and examine to what extent they can yield a
reasonable reionisation history and thermal state of the intergalactic
medium at redshifts around $z\sim 3$. This serves as an important
benchmark for our self-consistent methodology to simulate galaxy
formation and reionisation, and for future improvements through
accounting of other sources and other wavelength ranges. We find that
star formation alone is sufficient for reionising the Universe by
redshift $z\sim6$.  For a suitable choice of the escape fraction and
the heating efficiency, our models are approximately able to account
at the same time for the one-point function and the power spectrum of
the Lyman-$\alpha$ forest.  The radiation field has an important
impact on the star formation rate density in our simulations and
significantly lowers the gaseous and stellar fractions in low-mass
dark matter halos. Our results thus directly demonstrate the
importance of radiative feedback for galaxy formation. The spatial and
temporal importance of this effect can be studied accurately with the
modelling technique explored here, allowing more faithful simulations
of galaxy formation.

\end{abstract}


\section{Introduction}

In the standard cosmological model, the primordial gas recombines and
becomes neutral around redshift $z \simeq 1000$. However, the absence of
Gunn-Peterson troughs in the spectra of high redshift quasars up to $z
\leq 6$ \citep{White2003,Fan2006} suggests that hydrogen is highly
ionised at low redshift. Thus, there must be a period in the history of
the Universe when hydrogen became ionised again, but it is still an
open question when the process of this {\em cosmic reionisation}
started, how it proceeded in detail, and which sources of radiation were
primarily responsible for it.

An important observational clue about reionisation is provided by the
total electron-scattering optical depth to the last scattering surface
of the cosmic microwave background (CMB), found to be $\tau _{\rm es}
= 0.08785 \pm 0.00072$ by the latest WMAP7 data release
\citep{Larson2010}. This points to an early start and possibly
extended period of reionisation. Further observational information
about the history of hydrogen reionisation can be inferred through
different astrophysical phenomena. The Gunn-Peterson troughs in quasar
spectra are sensitive to small trace amounts of neutral hydrogen
during the late stages of reionisation ($z \sim 6$); Lyman-$\alpha$
emitting galaxies probe the intermediate stages of reionisation ($7 <
z < 15$); the 21-cm line background and gamma ray bursts (GRBs) probe
the early stages of reionisation, when the Universe was mostly neutral
($10 < z < 30$); and finally, the cosmic microwave background (CMB)
polarisation provides important data on the free electron column
density integrated over a large range of redshifts
\citep{Alvarez2006}. In the future, upcoming observations from new
radio telescopes such as LOFAR promise to be able to map out the epoch
of cosmic reionisation in unprecedented detail, making theoretical
studies of this process especially timely.

Population III stars are commonly believed to be the first significant
sources of photons for reionisation. They are very massive,
short-lived stars and form around redshift $z \sim 30$, or even
earlier \citep{Gao2007}. Another, in principle plausible candidate for
causing reionisation, are quasars. \citet{Madau1999} compare
luminosity and spacial density functions of quasars and star-forming
galaxies and show that quasars alone are unlikely to be the dominant
sources of photons for reionising the Universe.  On the other hand
other groups \citep{Trac2009, Thomas2009} predict in recent work that
quasars can produce enough ionising photons to be able to reionise the
Universe.

In many scenarios, reionisation is assumed to be driven
mainly by high-mass objects with mass $M > 10^9 \, M_{\odot}$
\citep{CF2000}. However, the dim but abundant low mass sources ($M <
10^9 \, M_{\odot}$) may still strongly influence the way in which the
ionised regions grow \citep{Sokasian2003, CF2007}.  These small galaxies
are preferentially found in low-density environments along the cosmic
web and may contain many metal-free Pop-III stars in the early
Universe. It has been estimated that these low-mass halos account for
about 80\% of the total ionising power at $z \sim 7$ \citep{CF2007}.
Disregarding them may lead to an overestimate of the number of
photons required to reionise the Universe \citep{Sokasian2003}.

Numerous theoretical studies have begun to investigate in detail the
characteristic scales and the topology of the reionisation process
through the use of numerical simulations \citep[e.g.][]{GO1997,
  Miralda2000, Gnedin2000, GA2001, Sokasian2001, Ciardi2003,
  Sokasian2004, Iliev2006, Zahn2007, Croft2008, Shin2008, Wise2008}. 
  However, due to the high
computational cost and complexity of the radiative transfer problem,
most simulations, with very few exceptions \citep{GO1997, Kohler2007,
  Shin2008, Wise2008}, have treated
reionisation through post-processing applied to static or separately
evolved gas density fields.  This neglects the fact that the radiation
field may exert important feedback effects on galaxy formation itself
\citep[e.g.][]{Iliev2005, Yoshida2007, Croft2008}. It is therefore an
important task to develop more accurate theoretical models based on
self-consistent simulations where the ionisation field is evolved
simultaneously with the growth of cosmic structures, a topic that we
address in this work.

In the literature, predictions based on simulations for the onset of
reionisation range from redshifts $z \sim 30-40$
\citep{Iliev2007,Wise2008} to $z \sim 15-20$ \citep{Norman1998,
  Abel2002}.  Reionisation then probably proceeds in an inhomogeneous
and patchy fashion \citep{Lidz2007, Iliev2007}, reflecting the
inhomogeneous density distribution of the large-scale structure. At
first, many isolated ionised bubbles are formed. They then grow in size
from $\sim 1\, \rm Mpc$ during the early stages of reionisation up to $>
10\, \rm Mpc$ in the late phases \citep{Furla2006, Iliev2006}. Around
redshift $z \sim 13$ \citep{Iliev2006}, $8 < z < 10$ \citep{Lee2007}, or
$z \sim 6$ \citep{GF2006}, the ionised regions overlap. 
These values are very uncertain and depend strongly on
the modelling details of reionisation and the parameters of the
underlying galaxy formation simulations. However, it is plausible that
the future use of more self-consistent simulation techniques should be
able to reduce the systematic modelling uncertainties.

In this work, we therefore use the new radiative transfer algorithm we
developed in a previous study \citep{Petkova2009} to carry out
high-resolution simulations of cosmic structure growth in the proper
cosmological context.  The approximation to radiative transfer employed,
the optically-thin variable Eddington tensor approach
\citep{GA2001}, is fast enough to allow coupled
radiative-hydrodynamic simulations of the galaxy formation process. At
the same time, the employed moment-based approximation to the radiative
transfer problem can be expected to be still reasonably accurate for the
reionisation problem. In particular, thanks to the photon-conserving
character of our implementation of radiative transfer and of the
chemical network, ionisation fronts are bound to propagate with the
right speed. The Lagrangian smoothed particle approach (SPH) we use
automatically adapts to the large dynamic range in density developing in
the galaxy formation problem. Combined with the fully adaptive
gravitational force solver implemented in {\small GADGET}, this yields a 
numerical scheme that is particularly well suited for the cosmic structure
formation problem.

For a first assessment of our new approach we study simulations of the
standard $\Lambda$CDM cosmology and treat star formation and supernova
feedback with the ISM sub-resolution model developed in
\cite{Springel2003}. For simplicity, we shall here only consider
ordinary star-formation regions as sources of ionising radiation. We are
especially interested in whether the star formation predicted by the
simulations results in a plausible reionisation history of the Universe,
and whether it at the same time yields a thermal and ionisation state of
the IGM at intermediate redshifts that is consistent with that probed by
observations of the Lyman-$\alpha$ forest. Finally, we are interested in
possible differences induced in galaxy formation due to the spatially
varying radiative feedback in the radiative transfer simulations,
especially in comparison with the much simpler and so far widely adopted
treatment where a spatially homogeneous UV background is externally
imposed.

This paper is structured as follows. We start in
Section~\ref{sec:methods} with a brief summary our methods for
simulating hydrogen reionisation. We then present our results in
Section~\ref{sec:results}, focusing on the history of reionisation in
Section~\ref{sec:history}, the Lyman-$\alpha$ forest in
Section~\ref{sec:Lyman-a} and the feedback from reionisation in
Section~\ref{sec:feedback}. We end with a discussion and our conclusions
in Section~\ref{sec:conclusion}.

\section{Simulating hydrogen reionisation} \label{sec:methods}

\subsection{Radiative transfer modelling}

For our work we use an updated version of the cosmological simulation
code {\small GADGET} \citep{gadget1, gadget2}, combined with the
radiative transfer (RT) implementation of \citet{Petkova2009}. The RT
equation is solved using a moment-based approach similar to the one
proposed by \citet{GA2001}. The resulting partial differential
equation essentially describes an anisotropic diffusion of the photon
density field $n_\gamma$,\footnote[1]{Differently from
  \citet{Petkova2009} we solve equation (\ref{eqn:RTmom}) for the photon
  density $n_\gamma$, rather than the photon over-density, with respect
  to the hydrogen density $\tilde{n}_\gamma = n_\gamma / n_{\rm H}$. }
\be \frac{\partial n_\gamma}{\partial t}= c \frac{\partial}{\partial
  x_j}\left( \frac{1}{\kappa}\frac{\partial n_\gamma h^{ij}}{\partial
  x_i} \right) - c\,\kappa\, n_\gamma + s_\gamma ,
\label{eqn:RTmom}
\ee where $c$ is the speed of light, $\kappa$ is the absorption
coefficient, $h^{ij}$ is the Eddington tensor and $s_\gamma$ is the
source function. The closure relation for this particular moment-based
method is obtained by approximating the Eddington tensor $h^{ij}$ as \be
h^{ij} = \frac{P^{ij}}{{\rm Tr}(P)} , \ee where $P^{ij}$ is the
radiation pressure tensor \be P^{ij}(x) \propto \int {\rm d}^3 x'
\rho_{\ast}(\vec{x}') \frac{(\vec{x}-\vec{x}')_i(\vec{x}-\vec{x}')_j}
    {(\vec{x}-\vec{x}')^4}.  \ee This estimate of the Eddington tensors
    is  carried out in the optically thin regime, giving the method
    its name.

\bfig
\begin{center}
\includegraphics[width=0.4\textwidth]{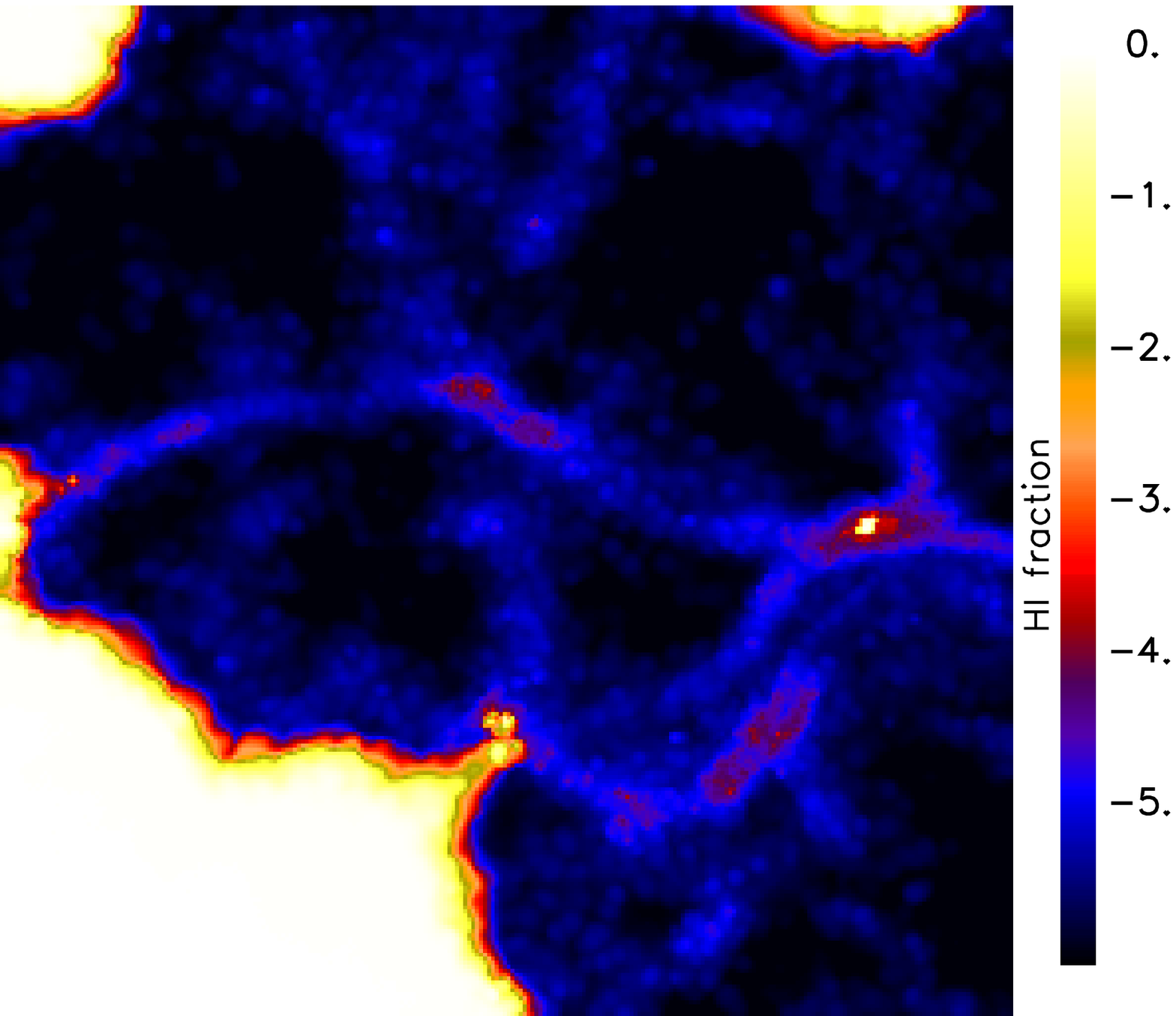}
\end{center}
\caption{Ionised fraction in a slice through the simulated volume at
  evolution time $t = 0.05\, \rm Myr$ of the cosmological density field
  test described in \citet{Iliev2006b} \label{fig:static}} 
\efig

The abundances of hydrogen and helium species are updated by accounting 
for the processes of photo-ionisation, collisional ionisation and recombination:
\be
  \frac{\partial \tilde n_{\rm HII}}{\partial t} = \gamma_{\rm
    HI} \tilde n_{\rm HI} \tilde n_e n_{\rm H} + c \sigma_0 \tilde
  n_{\rm HI} \tilde   n_\gamma n_{\rm H} - 
  \alpha_{\rm HII} \tilde n_{e} \tilde n_{\rm HII} n_{\rm H} ,
\ee

\bea
  \frac{\partial \tilde n_{\rm HeII}}{\partial t} = \gamma_{\rm
    HeI} \tilde n_{\rm HeI} \tilde n_e n_{\rm H} &-& \gamma_{\rm
    HeII} \tilde n_{\rm HeII} \tilde n_e n_{\rm H}\\ &+&
  \alpha_{\rm HeIII} \tilde n_{\rm HeIII} \tilde n_{e} n_{\rm
    H} ,
\eea

\be \frac{\partial \tilde n_{\rm HeIII}}{\partial t} = \gamma_{\rm HeII}
\tilde n_{\rm HeII} \tilde n_e n_{\rm H} - \alpha_{\rm HeIII} \tilde
n_{\rm HeIII} \tilde n_{e} n_{\rm H} , \ee where all abundances of
hydrogen species ($\tilde n_{\rm HI}$, $\tilde n_{\rm HII}$) are expressed
in dimensionless form relative to the total number density $n_{\rm H}$
of hydrogen, all helium abundances are fractions with respect to the
helium number density, and the electron abundance is expressed as
$\tilde n_e = n_e / n_{\rm H}$. Furthermore, $\gamma$ denotes the
collisional ionisation coefficient, $\alpha$ is the recombination
coefficient and $\sigma_0 = 6.3\times 10^{-18} \, \rm cm^2$ is the
photo-ionisation cross-section for hydrogen at $13.6\,\rm eV$. These
rate equations are solved using a semi-implicit scheme.

\bfig
\begin{center}
\includegraphics[width=0.45\textwidth]{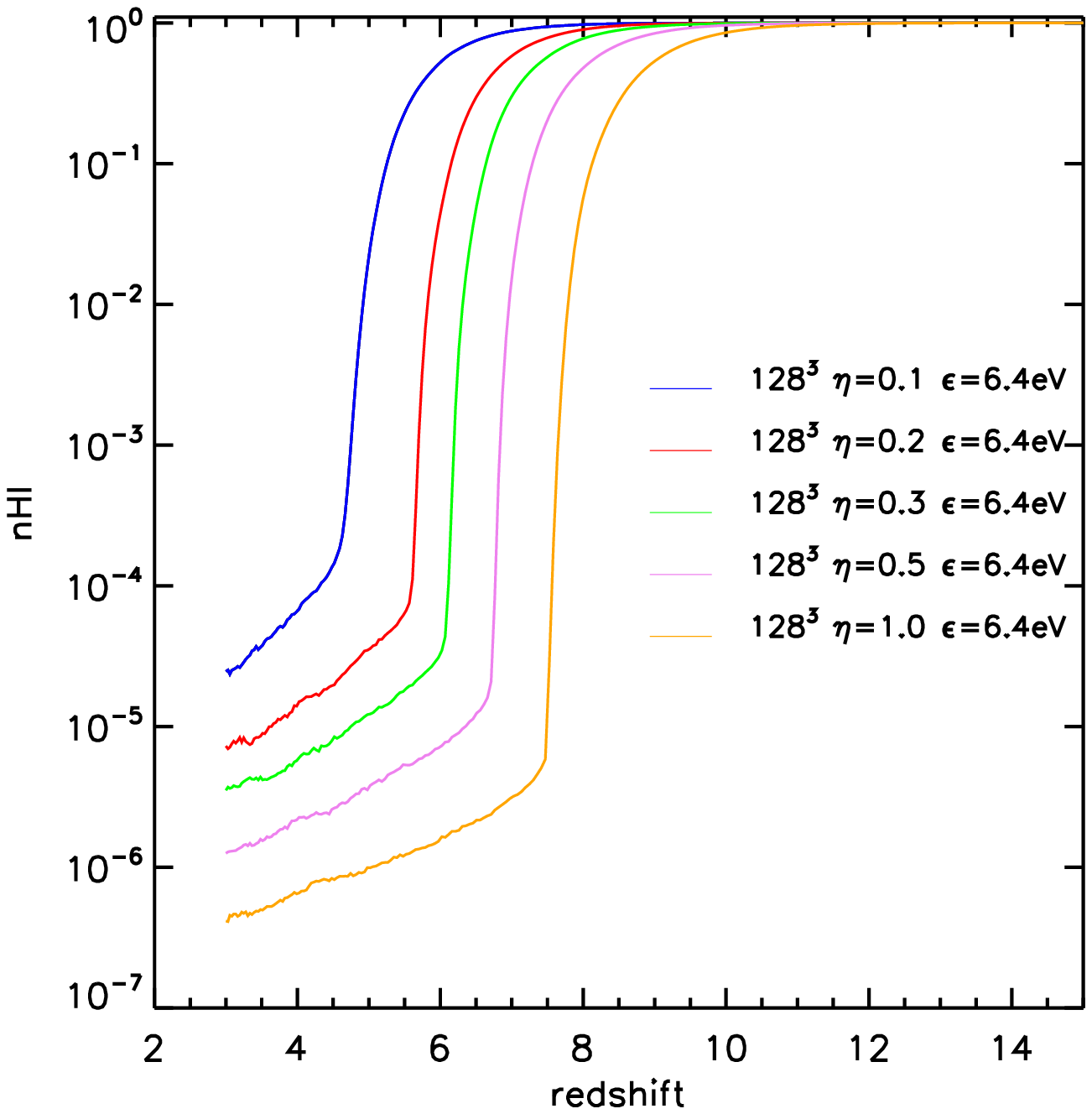}
\end{center}
\caption{Volume averaged neutral fraction as a function of redshift
  for the low resolution simulations. The
  different colours represent simulation results for different values of
  the efficiency parameter (or `ISM escape fraction') $\eta = 0.1,\,
  0.2,\, 0.3,\, 0.5,\, 1.0$. Reionisation happens earlier for higher
  efficiency since then more photons become available for ionising the
  gas. In all cases, the final phase of reionisation proceeds rapidly;
  over a small range of redshift, the neutral volume fraction drops from
  10\% to negligibly small values.\label{fig:nHI}} \efig

\bfig
\begin{center}
\includegraphics[width=0.45\textwidth]{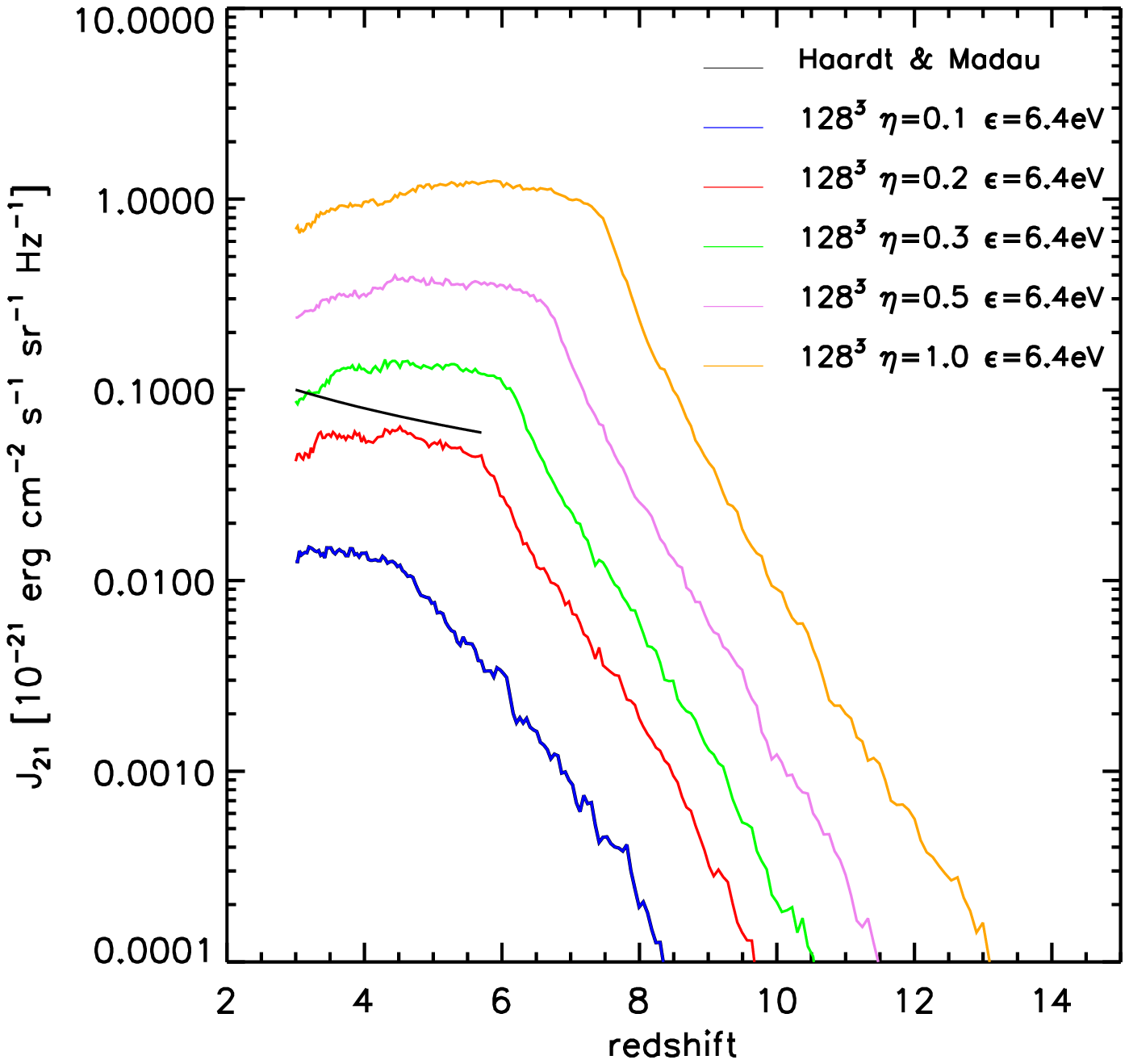}
\end{center}
\caption{Ionising background for the low resolution simulation 
  as a function of redshift for different
  efficiency $\eta = 0.1,\, 0.2,\, 0.3,\, 0.5,\, 1.0$. As expected, the
  background becomes higher for a larger efficiency since more of the
  photons emitted by the sources are made available to build up the
  ionising background. The thin solid line shows the background
  computed by \citet{Haardt1996} for quasars, which provides for an
  interesting comparison. Our results for the time evolution of the
  neutral volume fraction agree quite well with previous studies
  \citep[e.g.][]{Gnedin2000}.\label{fig:j21}} \efig

The photo-heating of the gas by the ionising part of the
spectrum is described by the heating rate 
\bea
\Gamma &=& n_{\rm HI} \int_{\nu_0}^\infty {\rm d \nu} \frac{4\pi
  I_\nu}{h\nu} \sigma_\nu (h\nu - h\nu_0)\\
&=&n_{\rm HI}\,c\,n_\gamma\,\tilde{\epsilon}_{\rm HI}\, \tilde{\sigma}_{\rm
  HI},
\eea
where
\be
  \tilde \sigma = \int_{\nu_0}^\infty {\rm d\nu} \frac{4\pi I_\nu}
    {h\nu}\sigma_\nu \times \left( \int_{\nu_0}^\infty {\rm d\nu}
  \frac{4\pi I_\nu}
  {h\nu}\right)^{-1}
\ee
is a frequency averaged photo-ionisation cross-section and
\be
  \tilde \epsilon = \int_{\nu_0}^\infty {\rm d\nu} \frac{4\pi
    I_\nu
    \sigma_\nu}{h\nu}(h\nu - h\nu_0) \times \left( \int_{\nu_0}^\infty {\rm d\nu}
  \frac{4\pi I_\nu \sigma_\nu}
  {h\nu}\right)^{-1}
  \label{eqn:epsilon}
\ee
is a frequency averaged photon excess energy \citep{Spitzer1998}. For
our calculations, we usually assume a black body spectrum with $T_{\rm
  eff}=10^5\, \rm K$, which leads to $\tilde \sigma_{\rm HI} = 1.63 \times
10^{-18} \, \rm cm^2$ and $\tilde \epsilon_{\rm HI} = 6.4\, \rm
eV$. However, given the uncertainties in accurately calculating the net
energy input during the photo-ionisation transition, which involves
non-equilibrium processes, we will later on vary $\tilde \epsilon_{\rm
  HI}$ in order to investigate the influence of different heating
efficiencies.

We have tested our radiative transfer implementation on the static
cosmological density field that was used in the radiative transfer code
comparison study by \citet{Iliev2006b}. In Figure \ref{fig:static} we
show the ionised fraction in a slice through the simulated volume at
evolution time $t=0.05\, \rm Myr$. Reassuringly, our result is in good
agreement with the ones obtained by other radiative codes in the
comparison study.

\subsection{Treatment of star formation and radiative cooling}

We follow the radiative cooling of helium and hydrogen with a
non-equilibrium treatment, with the rates as cited in
\citet{Petkova2009}. Winds and star formation in the dense, cold gas is modelled
in a sub-resolution fashion \citep{Springel2003}, where the
interstellar medium is pictured as being composed of cold clouds that
are embedded at pressure equilibrium in a hot tenuous phase that is
heated by supernova explosions. Through the evaporation of clouds,
this establishes a tight self-regulation of the star formation
rate. Previous work has shown that this model converges well with
numerical resolution and yields star formation rates that are
consistent with the Kennicutt relation \citep{Kennicutt1998} observed
at low redshift. We shall here assume that the same star formation law
also holds at high redshift.

We use the star-forming regions in all simulated galaxies as sources of
ionising photons in our radiative transfer model.  We adopt an ionising
source luminosity of $\dot N_{\rm SFR} = 10^{53} \, \rm photons \,
M_{\odot}^{-1} \, yr$ \citep{Madau1999}, which relates the number of
emitted photons to the star formation rate in units of ${\rm M}_{\odot}
\rm yr^{-1}$. This source luminosity is released individually by every
star-forming gas particle, hence the number of numerically represented
sources is a non-negligible fraction of all simulation
particles. Fortunately, the speed of our radiative transfer algorithm is
almost insensitive to the total number of sources, because the Eddington
tensor calculation can be carried out with a tree algorithm similar to
the gravity calculation. This insensitivity of the computational cost to
the number of sources is a significant advantage of the method used
here, and is not shared by most alternatives for treating radiative
transfer.

\bfigs
\includegraphics[width=0.29\textwidth]{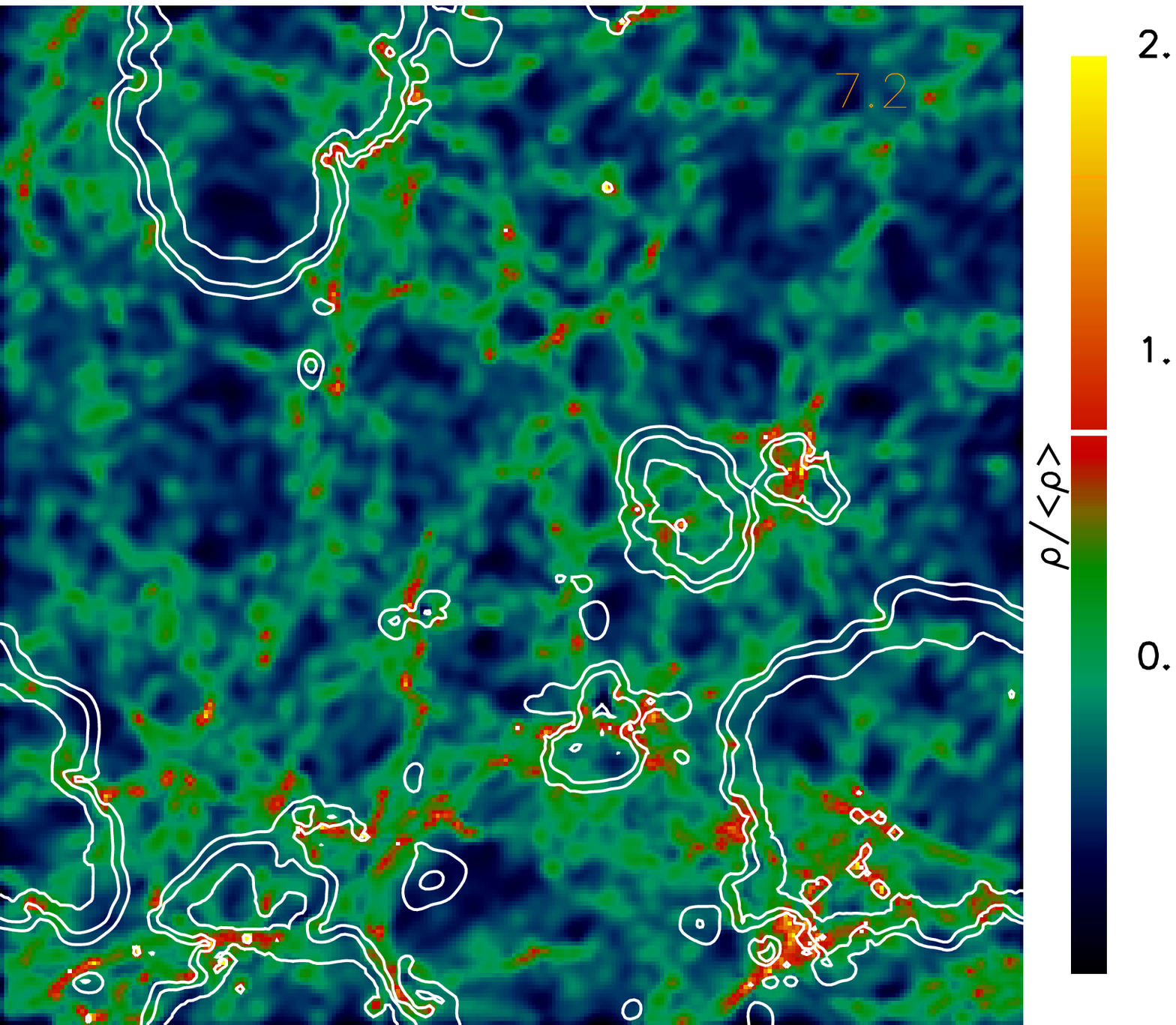}
\includegraphics[width=0.29\textwidth]{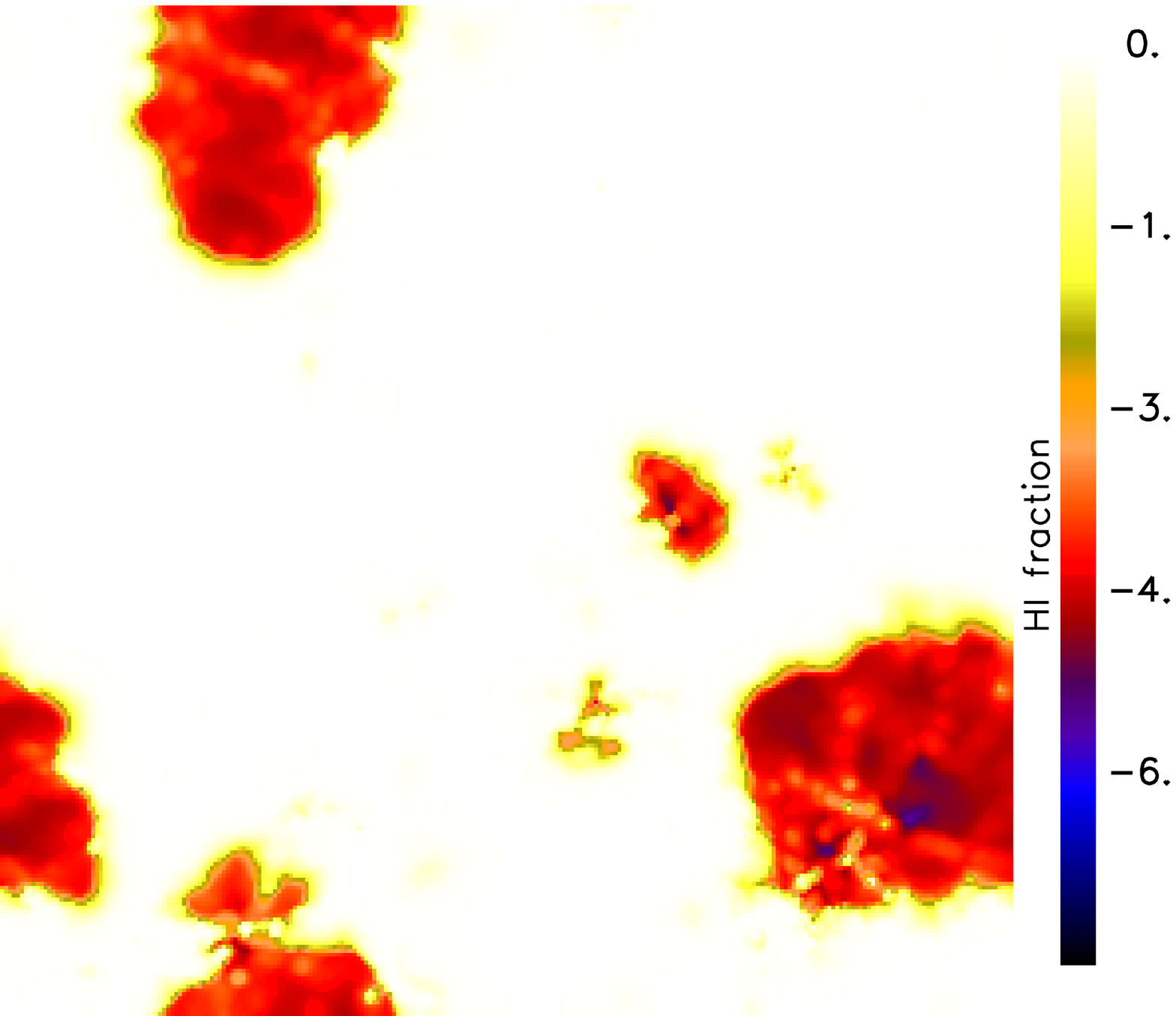}
\includegraphics[width=0.29\textwidth]{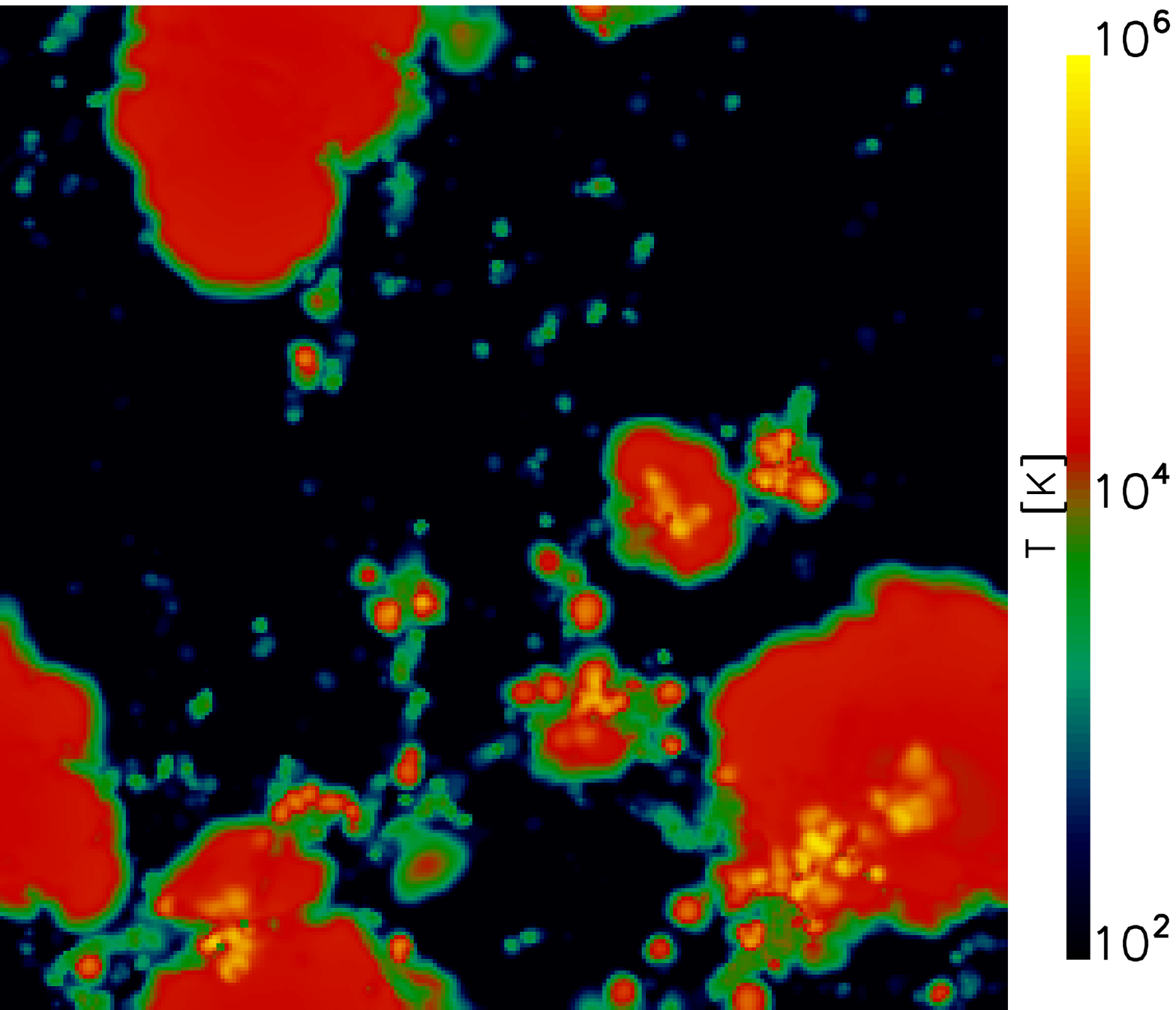}
\includegraphics[width=0.29\textwidth]{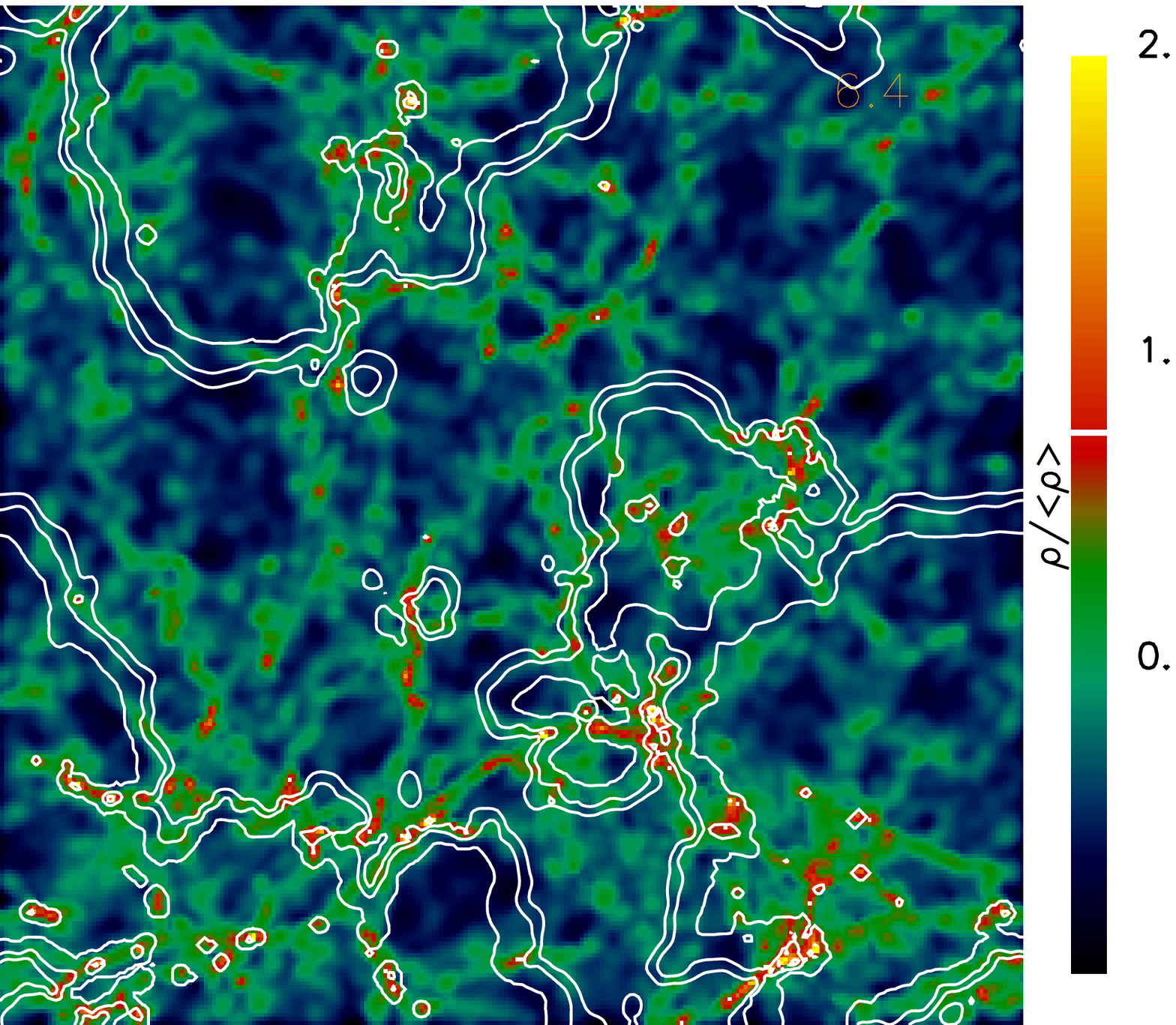}
\includegraphics[width=0.29\textwidth]{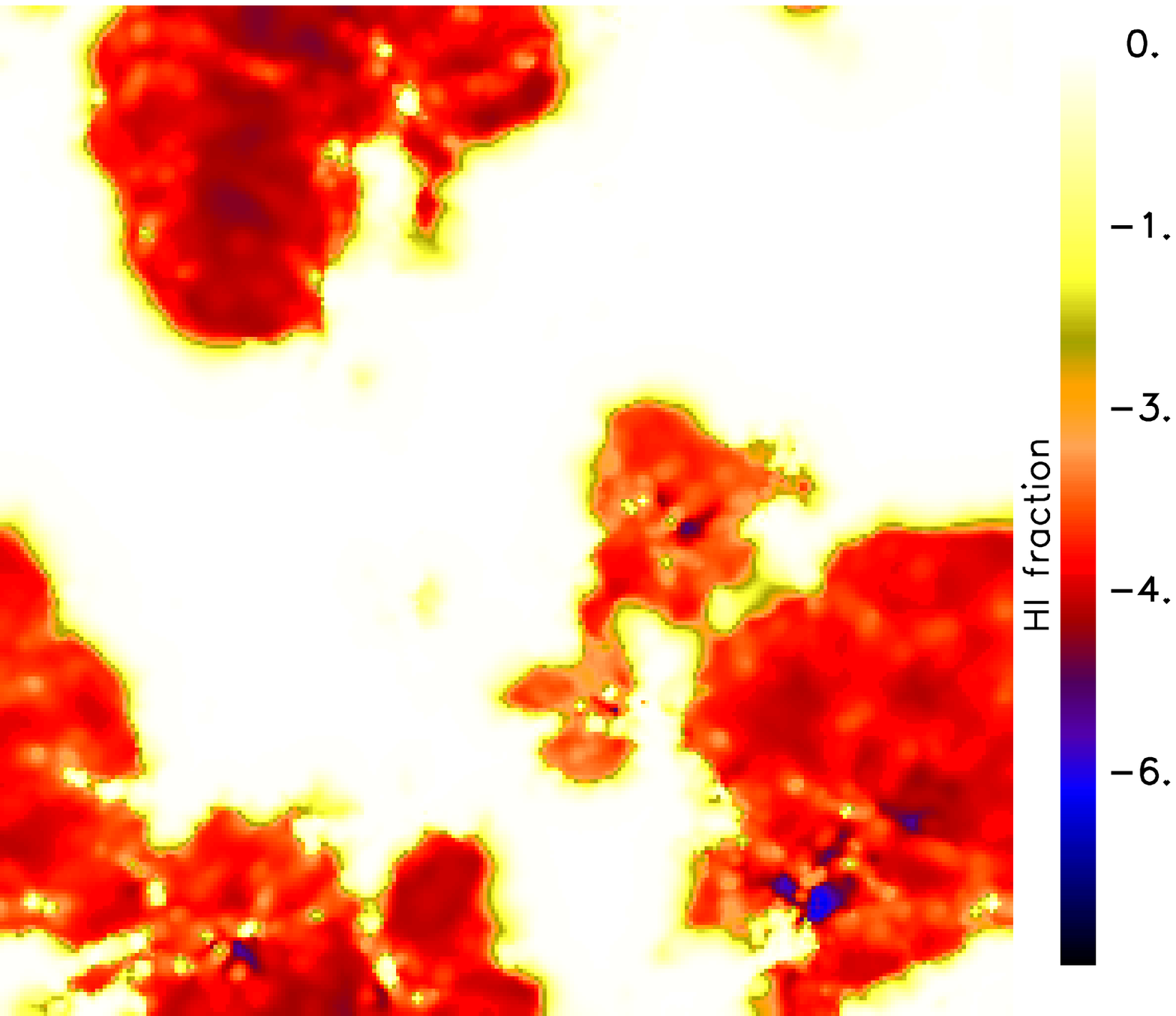}
\includegraphics[width=0.29\textwidth]{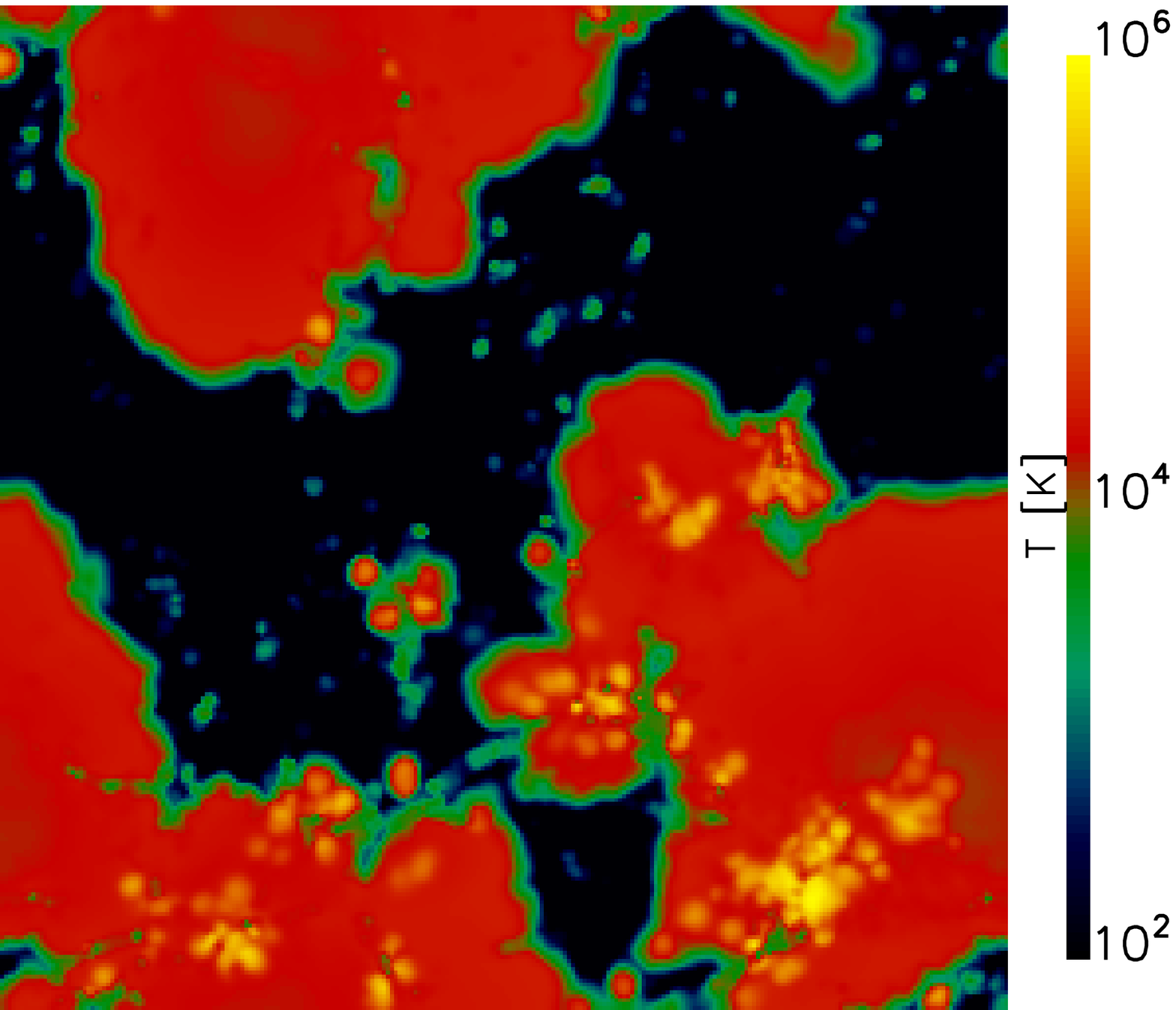}
\includegraphics[width=0.29\textwidth]{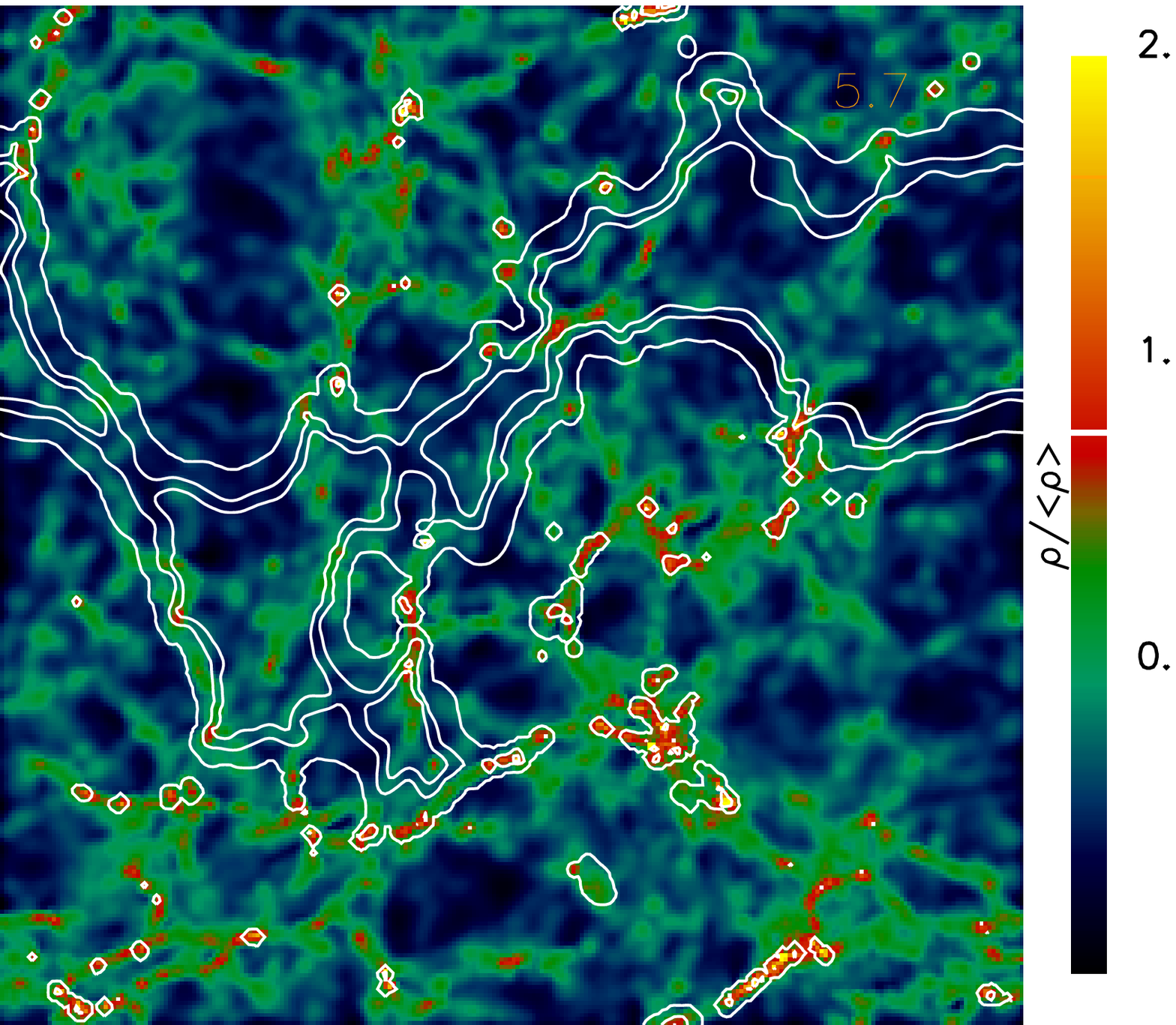}
\includegraphics[width=0.29\textwidth]{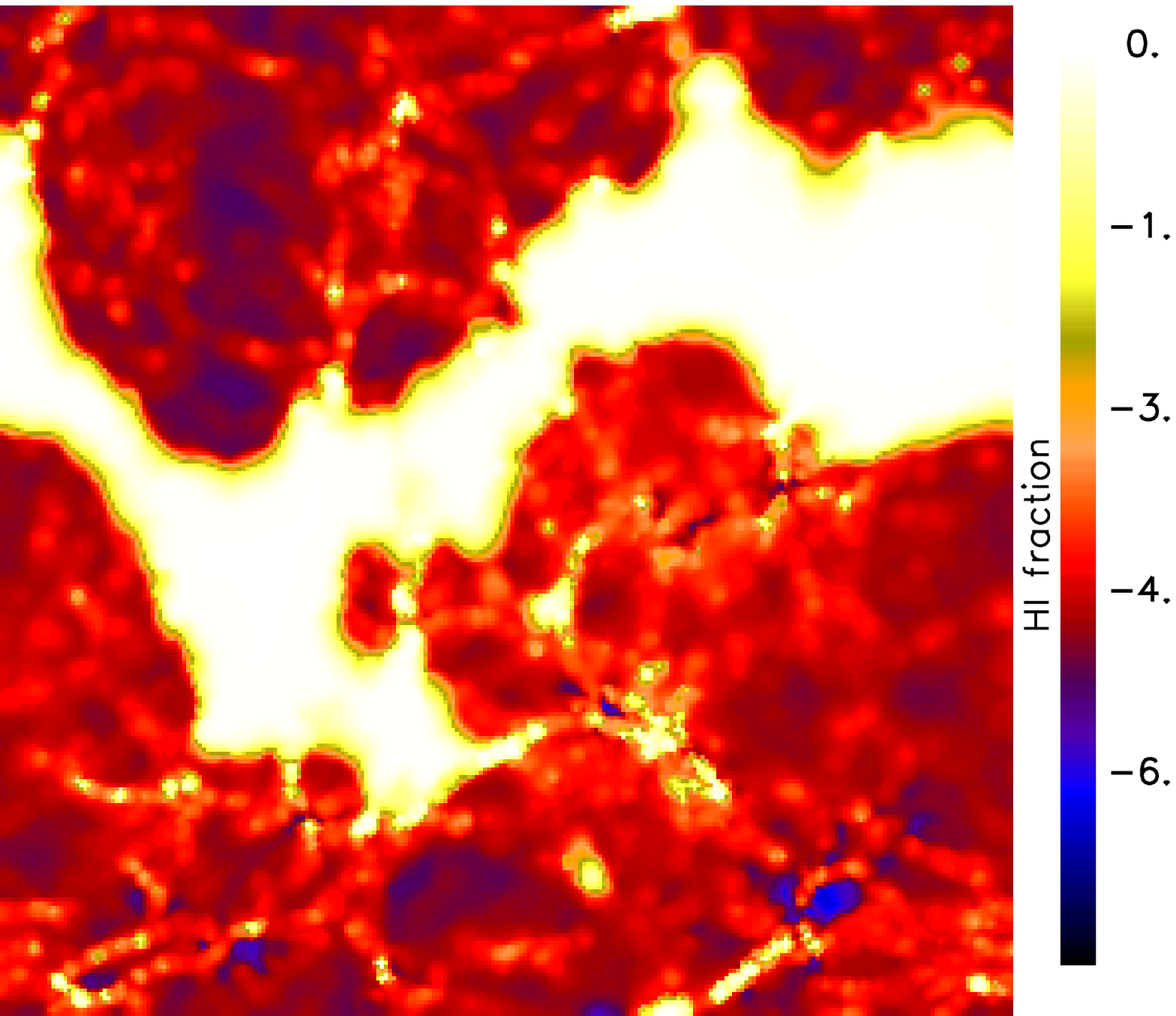}
\includegraphics[width=0.29\textwidth]{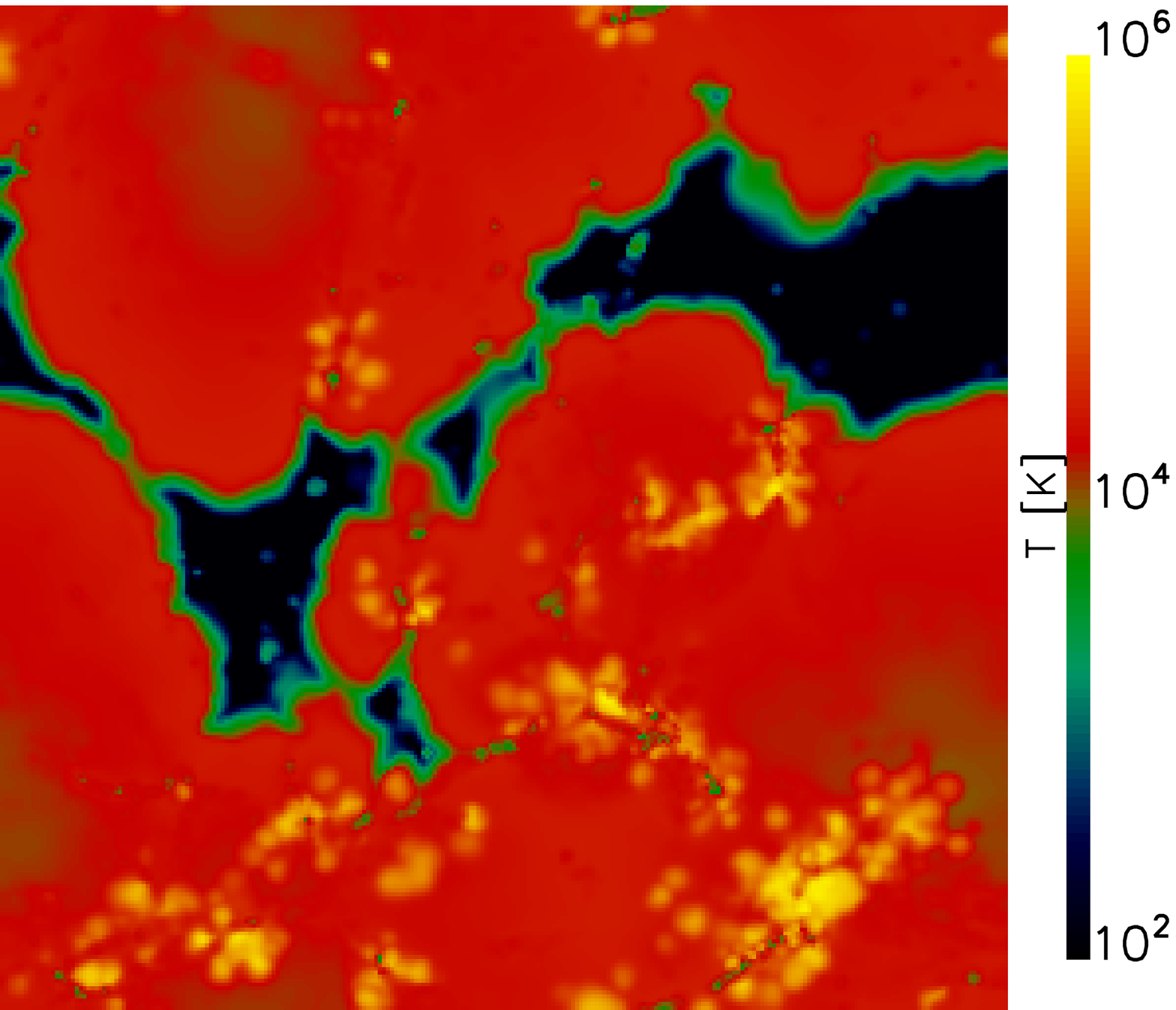}
\includegraphics[width=0.29\textwidth]{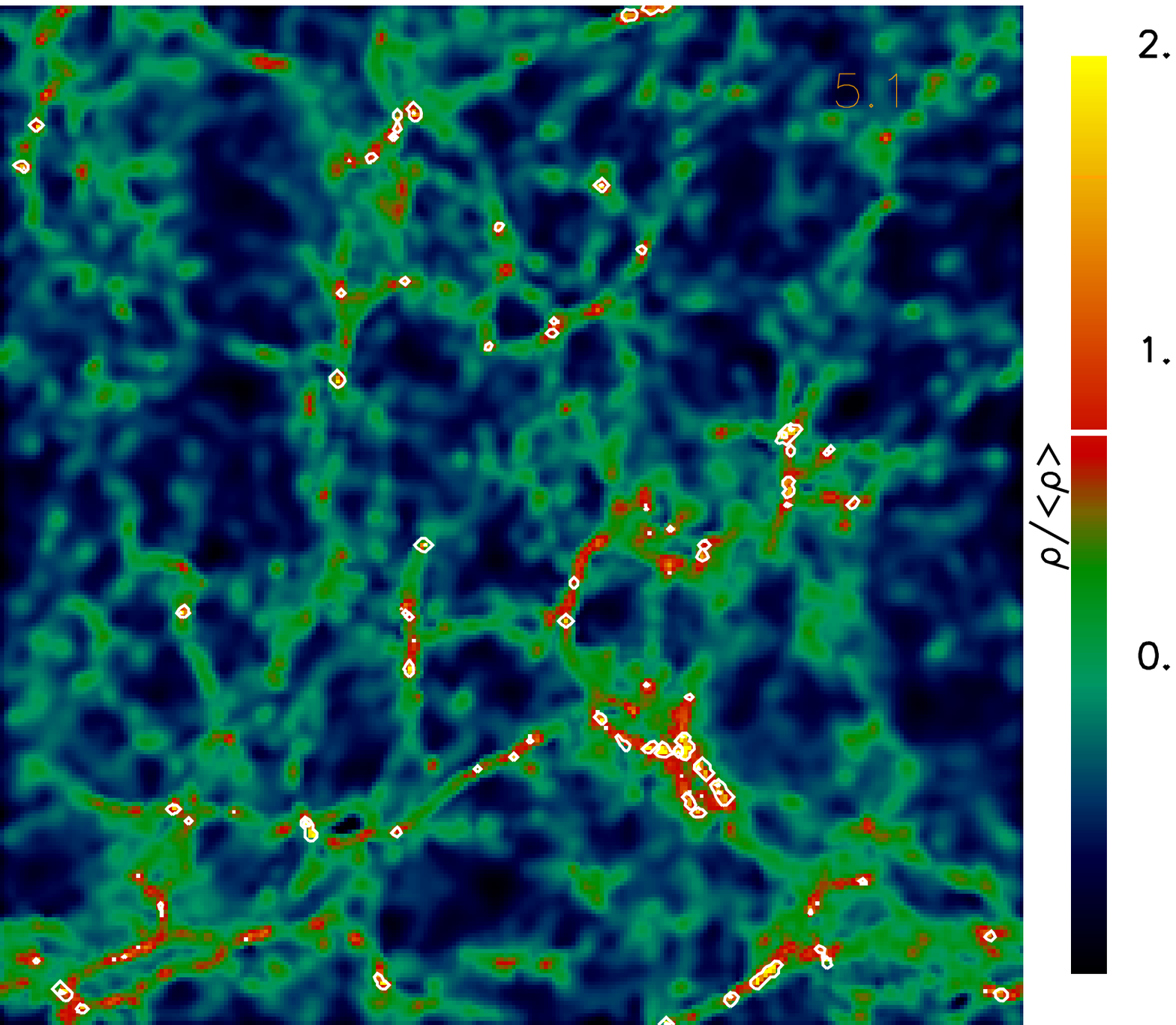}
\includegraphics[width=0.29\textwidth]{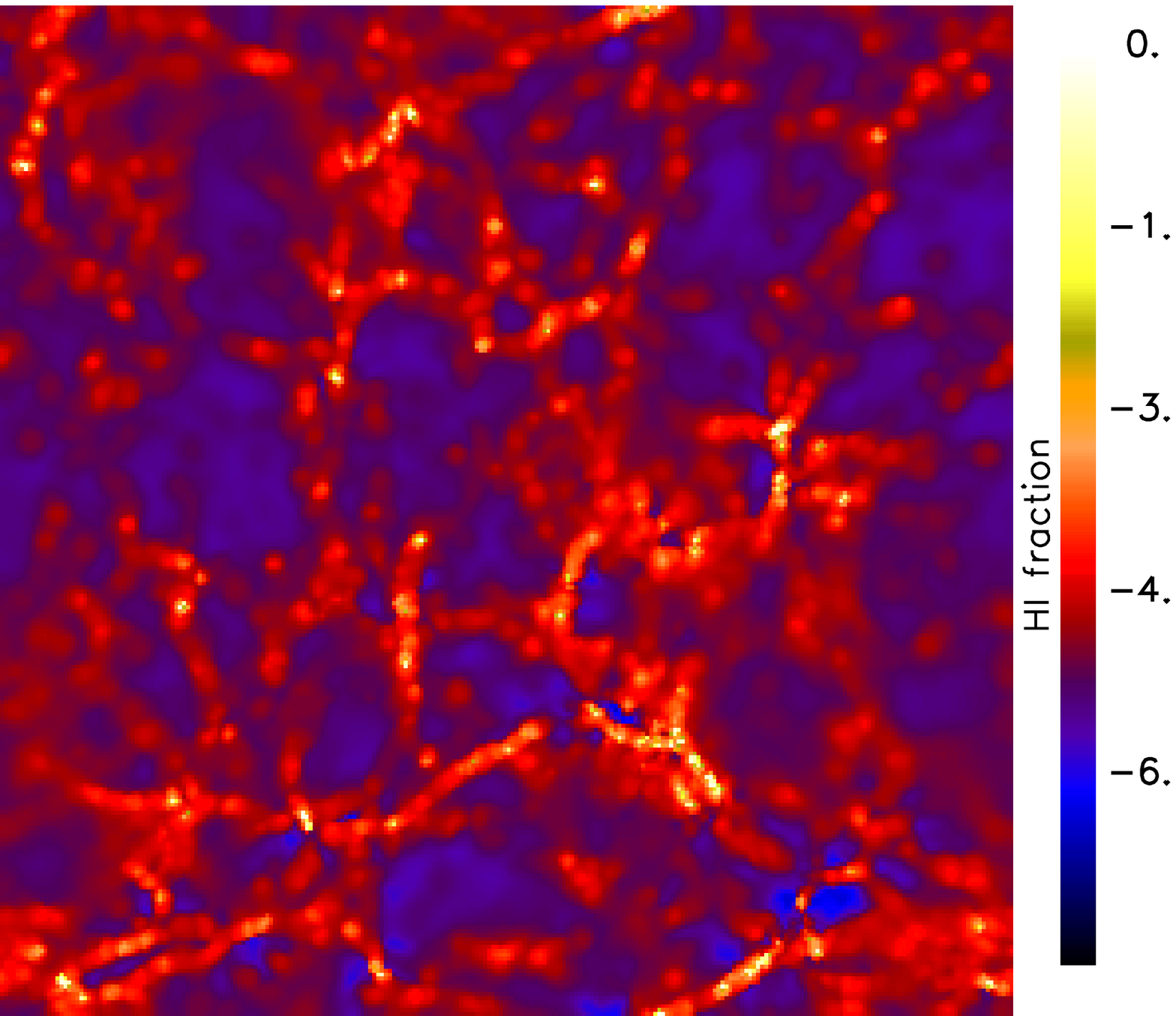}
\includegraphics[width=0.29\textwidth]{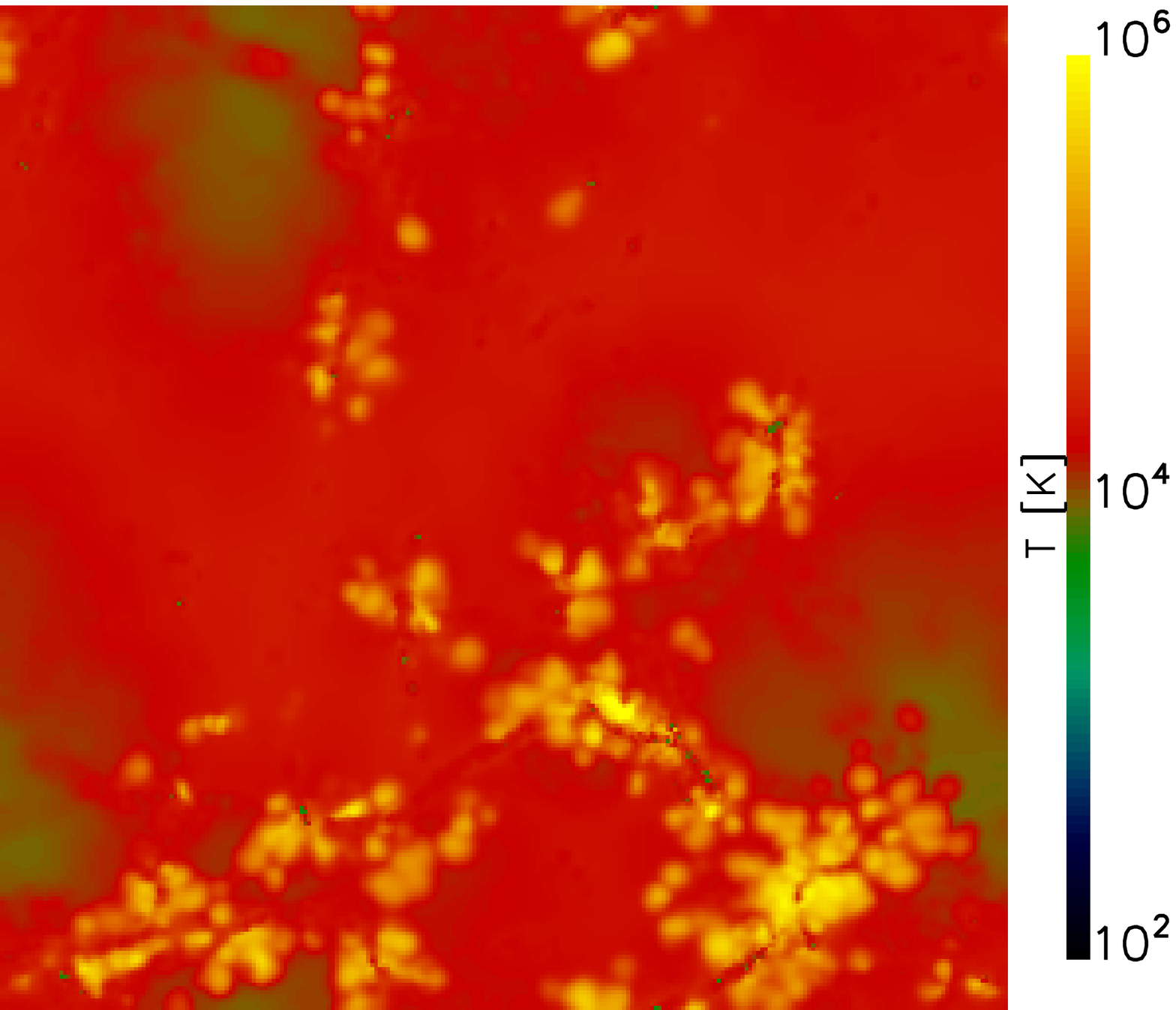}
\includegraphics[width=0.29\textwidth]{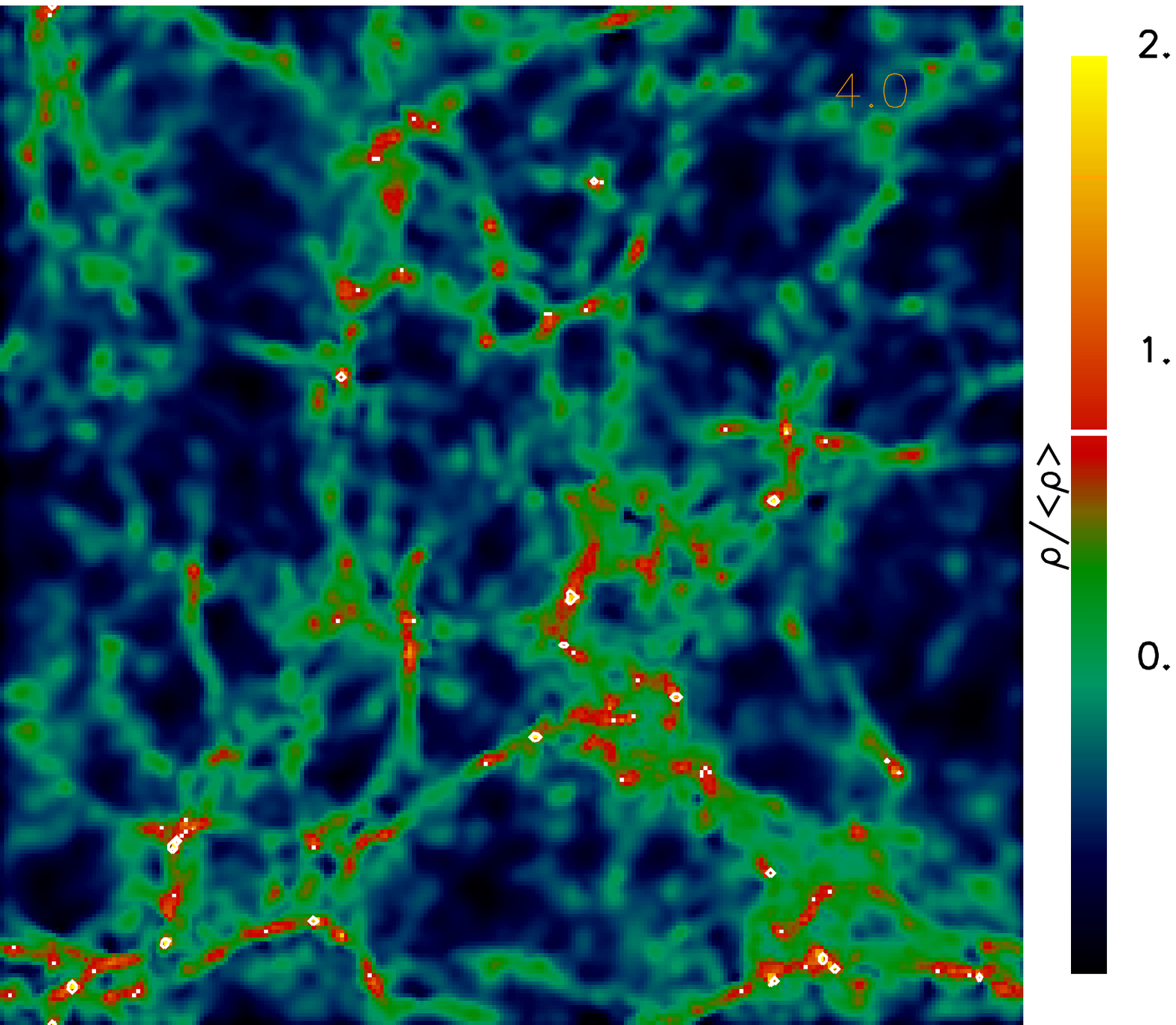}
\includegraphics[width=0.29\textwidth]{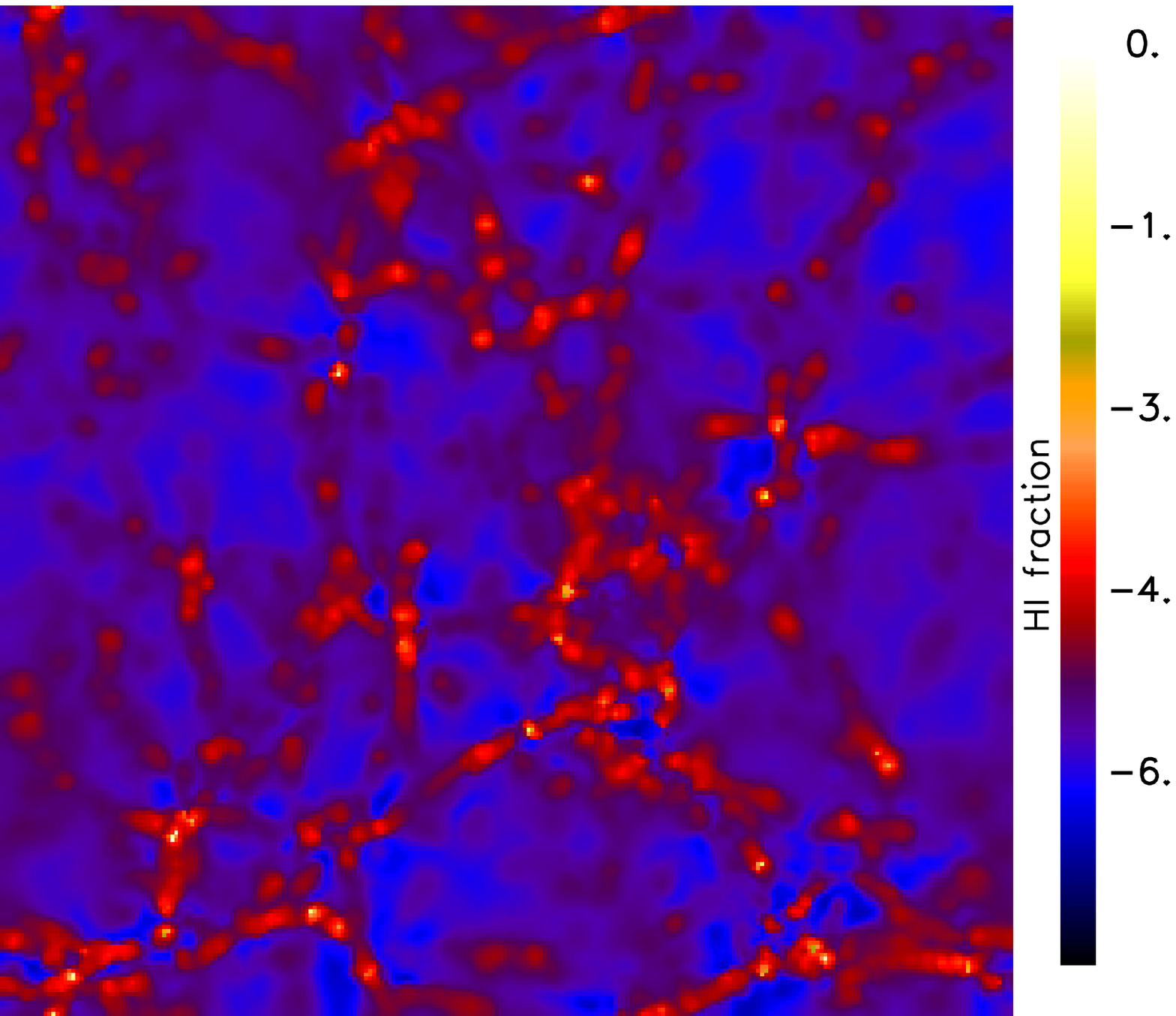}
\includegraphics[width=0.29\textwidth]{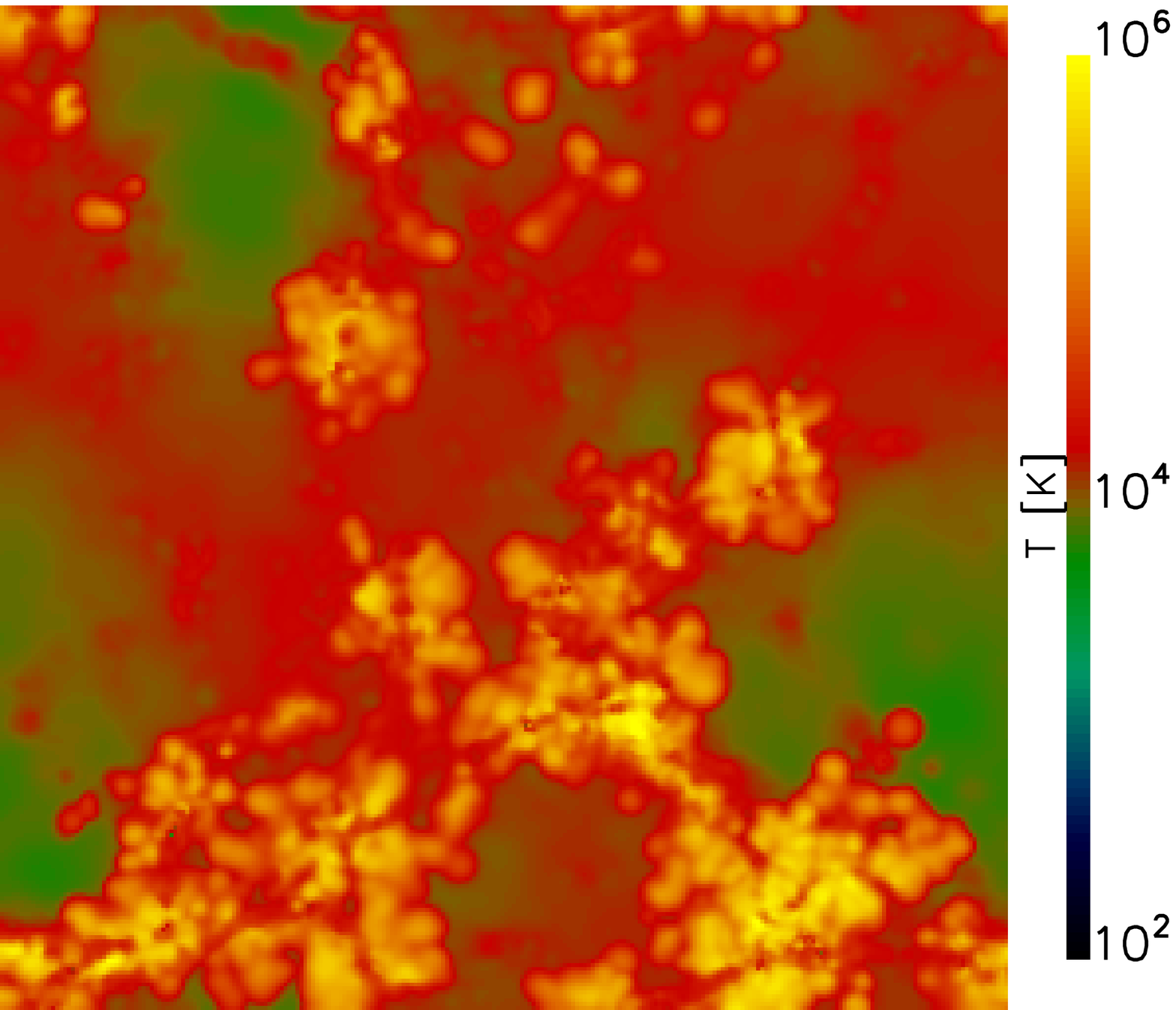}
\caption{Slices through the middle of the simulated volume for the low
  resolution simulation realisation with $\eta = 0.1$ and $\tilde
  \epsilon = 30 \, \rm eV$. The maps are showing the density (left)
  with contours at ionised fraction $\tilde n_{\rm HII} = 0.001$,
  $0.5$ and $0.9$ overlaid, ionised fraction (middle), and temperature
  (right) of the gas. The snapshots show a time evolution from top to
  bottom, with individual redshifts $z=7.2$, $z=6.4$, $z=5.7$, $z=5$
  and $z=4$, respectively.  \label{fig:dens_evolve}} \efigs

To account for the uncertain absorption that occurs in reality in the
spatially unresolved multi-phase structure of our simulations, we impose
a phenomenological efficiency factor $\eta$ on the source
luminosities. In our simulation set we explore values in the range $\eta
= 0.1-1.0$ to get a feeling for the sensitivity of our results to this
uncertainty. We note however that $\eta$ should not be confused with
what is usually called galaxy escape fraction, which has a slightly
different meaning. Our $\eta$ is meant to be just an `interstellar
medium escape fraction' whereas photon losses in the gaseous halos of
galaxies will be taken into account self-consistently in our
simulations.

\bfigs
\begin{center}
\includegraphics[width=0.47\textwidth]{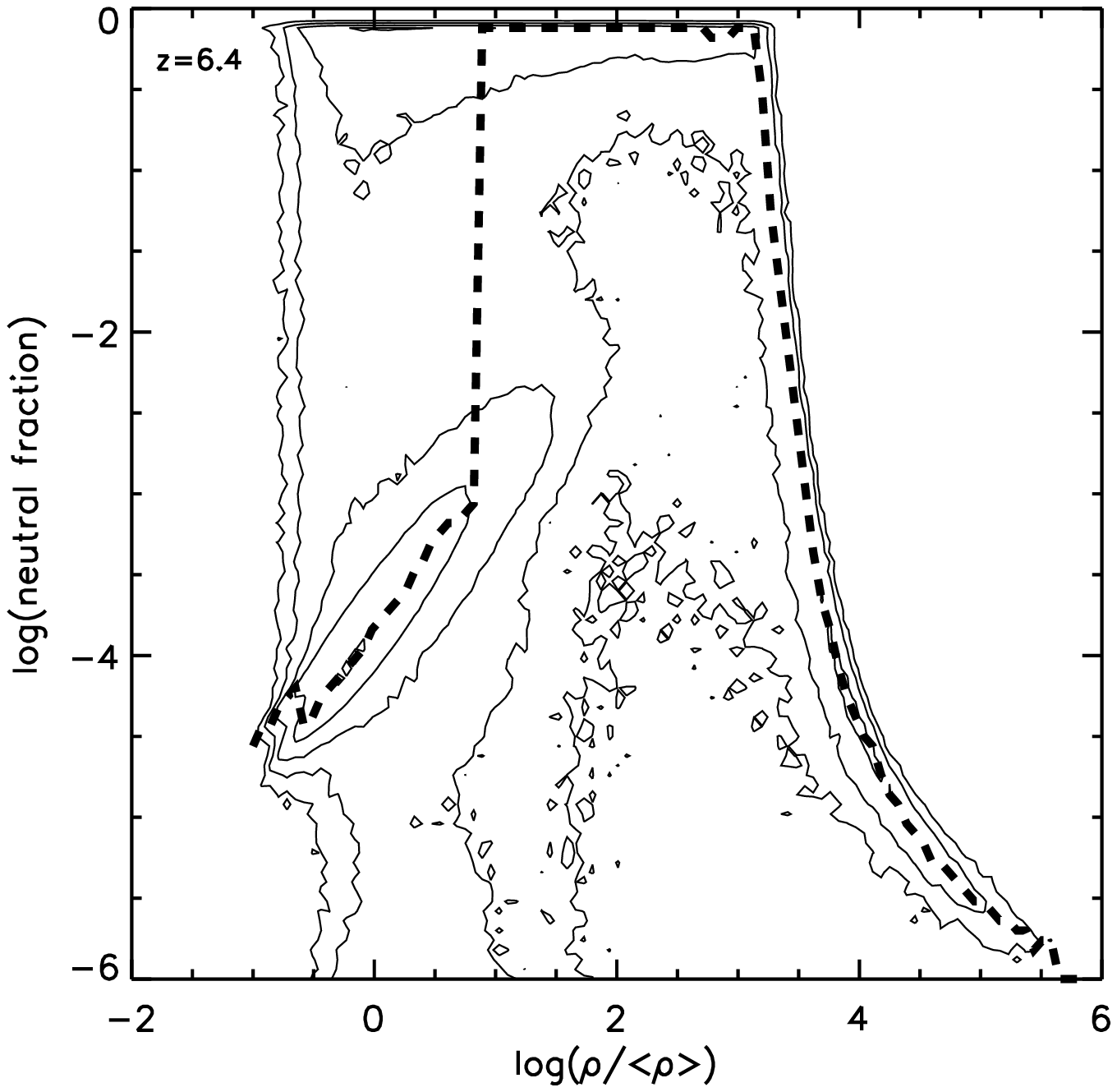}
\includegraphics[width=0.47\textwidth]{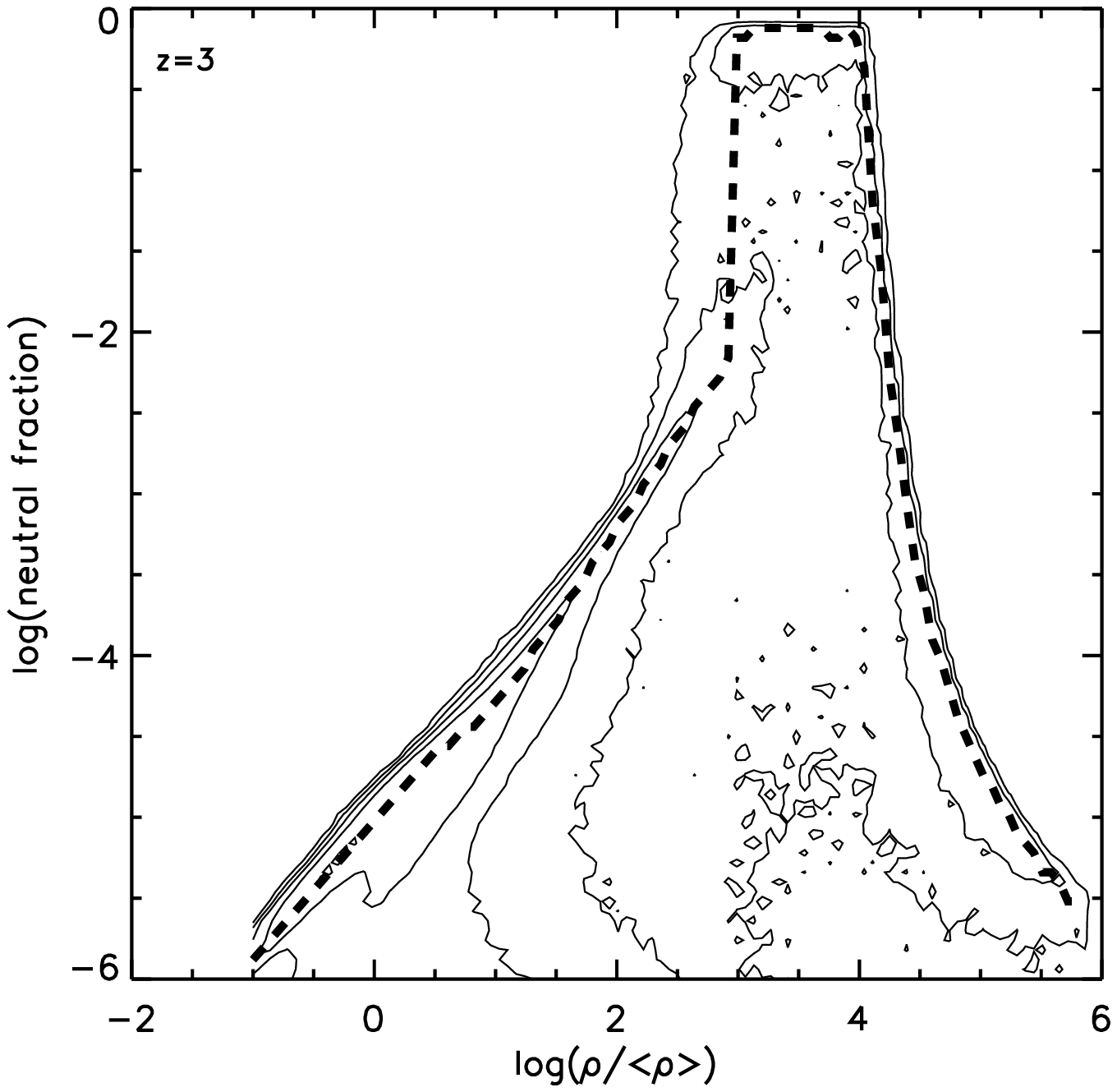}
\end{center}
\caption{Scatter plot of the neutral fraction as a function of
  over-density for two different redshift of the low resolution
  simulation realisation with $\eta = 0.1$ and $\tilde \epsilon =
  30 \, \rm eV$. The thick dashed line is the median neutral
  fraction. The neutral fraction shows a
  clear dependence with density, where high density regions are more
  ionised than low density regions. Very high density gas is highly
  ionised as well, because it is in the star forming
  stage. \label{fig:q_rho}} 
\efigs

\bfigs
\begin{center}
\includegraphics[width=0.48\textwidth]{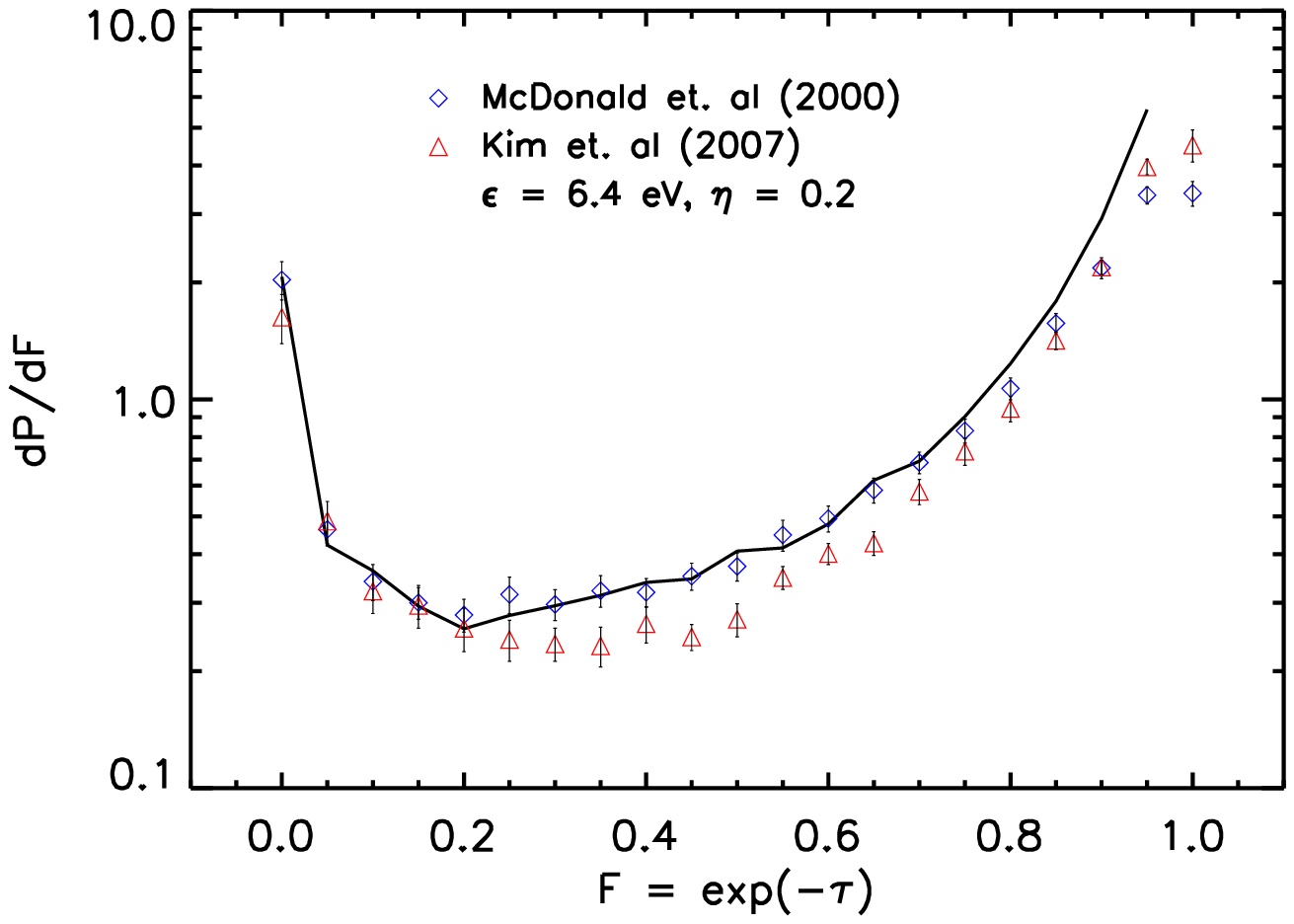}
\includegraphics[width=0.48\textwidth]{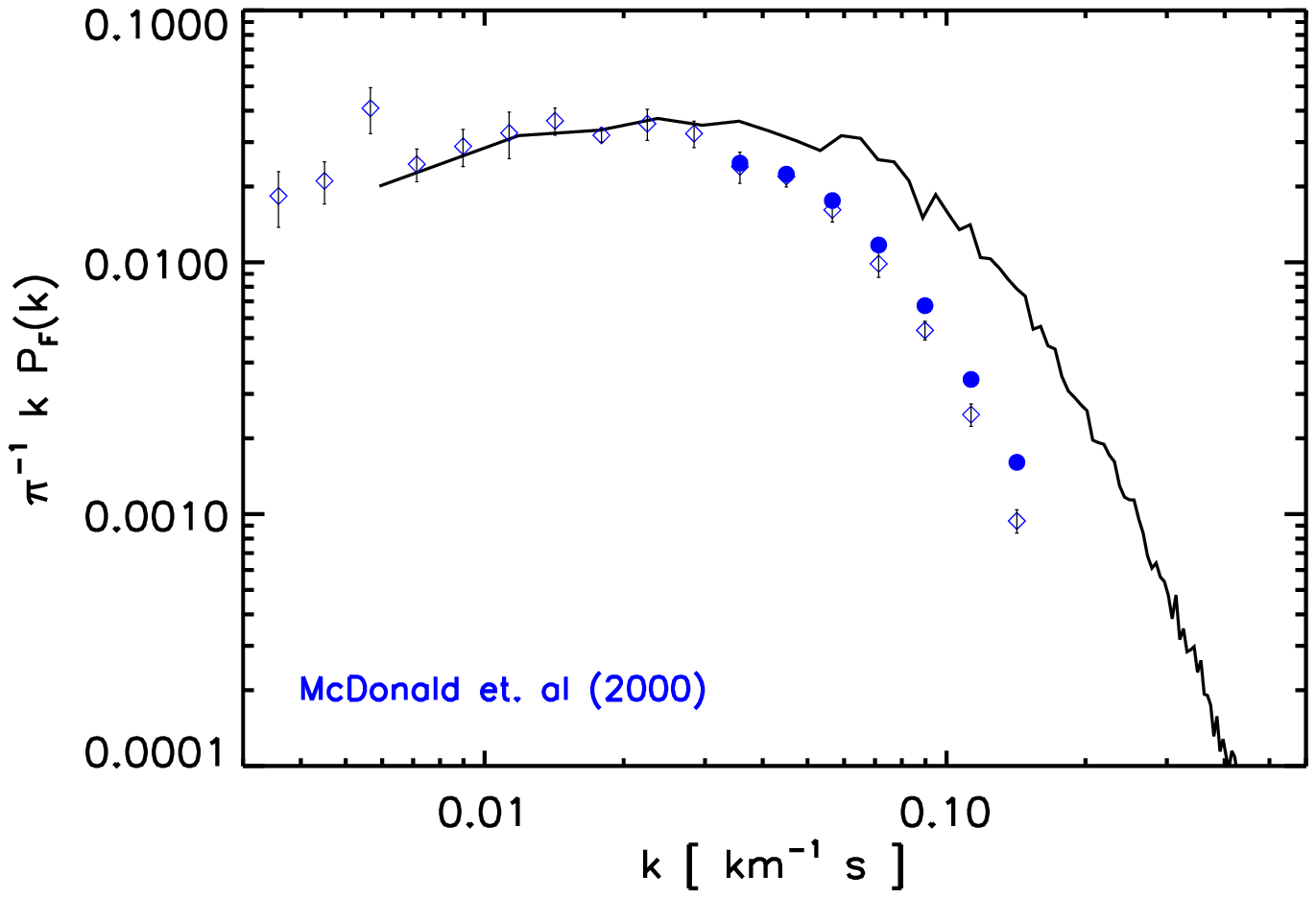}
\end{center}
\caption{Lyman-$\alpha$ flux probability (left) and power spectrum
  (right) from the low resolution
  simulation with $\eta = 0.2$ and $\tilde \epsilon = 6.4 \, \rm eV$ at redshift $z=3$. The result
  is compared with observational data from \citet{McDonald2000} and
  \citet{Kim2007}. The
  flux probability agrees reasonably well with observations, whereas the
  power spectrum deviates at the high $k$ end. We discuss this problem in
  the text. \label{fig:flux_020}}
\efigs

\bfigs
\includegraphics[width=0.44\textwidth]{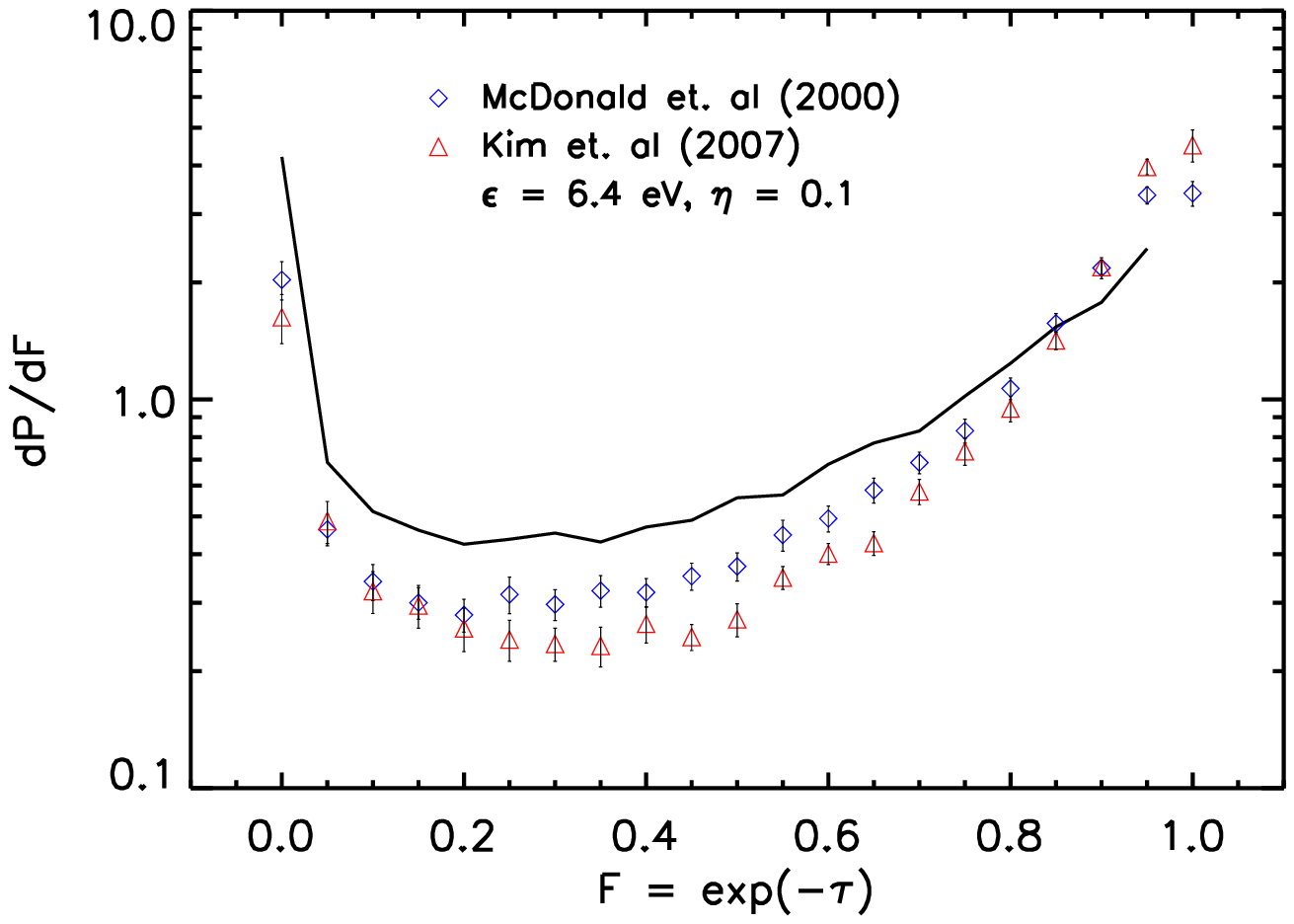}
\includegraphics[width=0.44\textwidth]{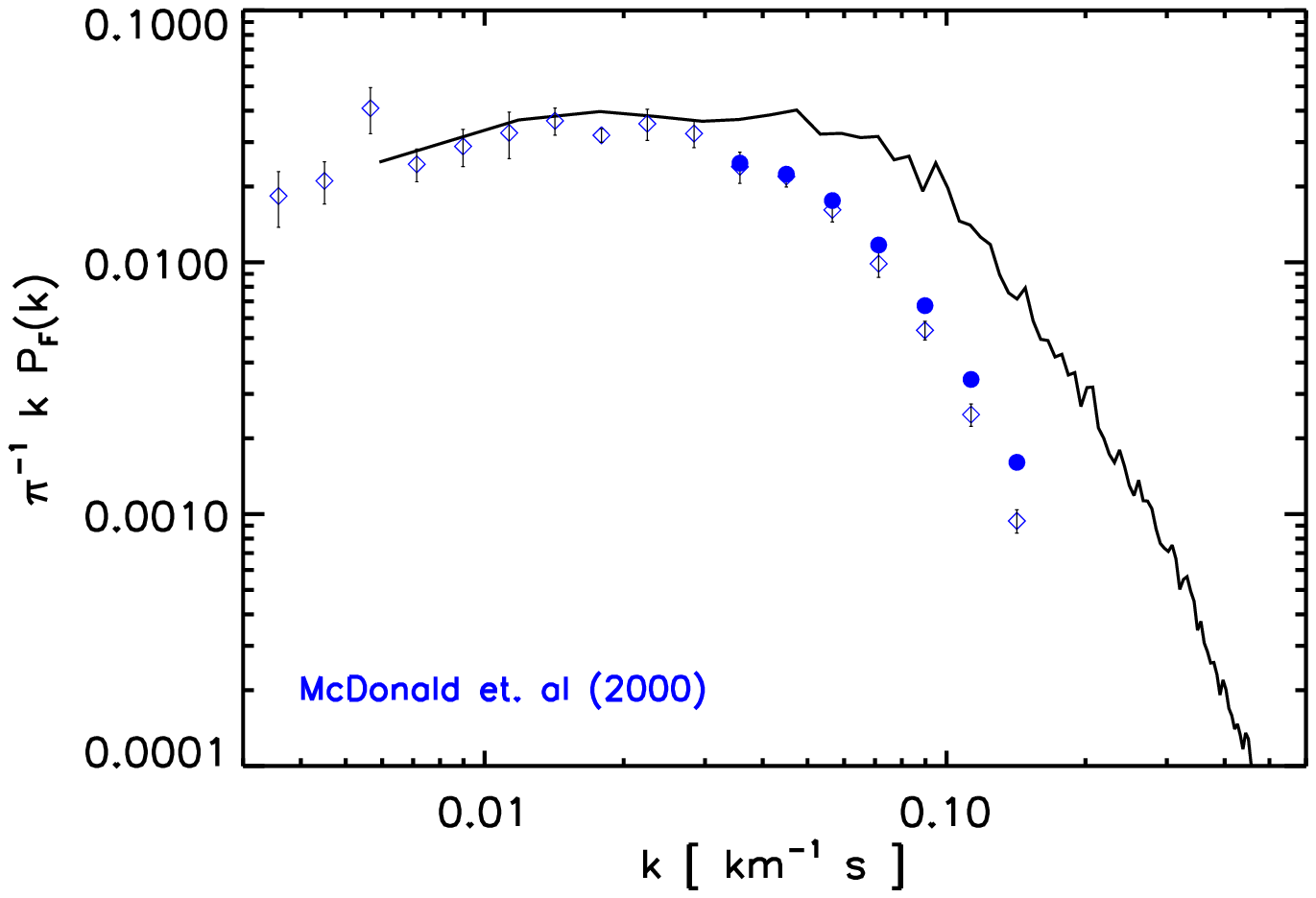}\\
\includegraphics[width=0.44\textwidth]{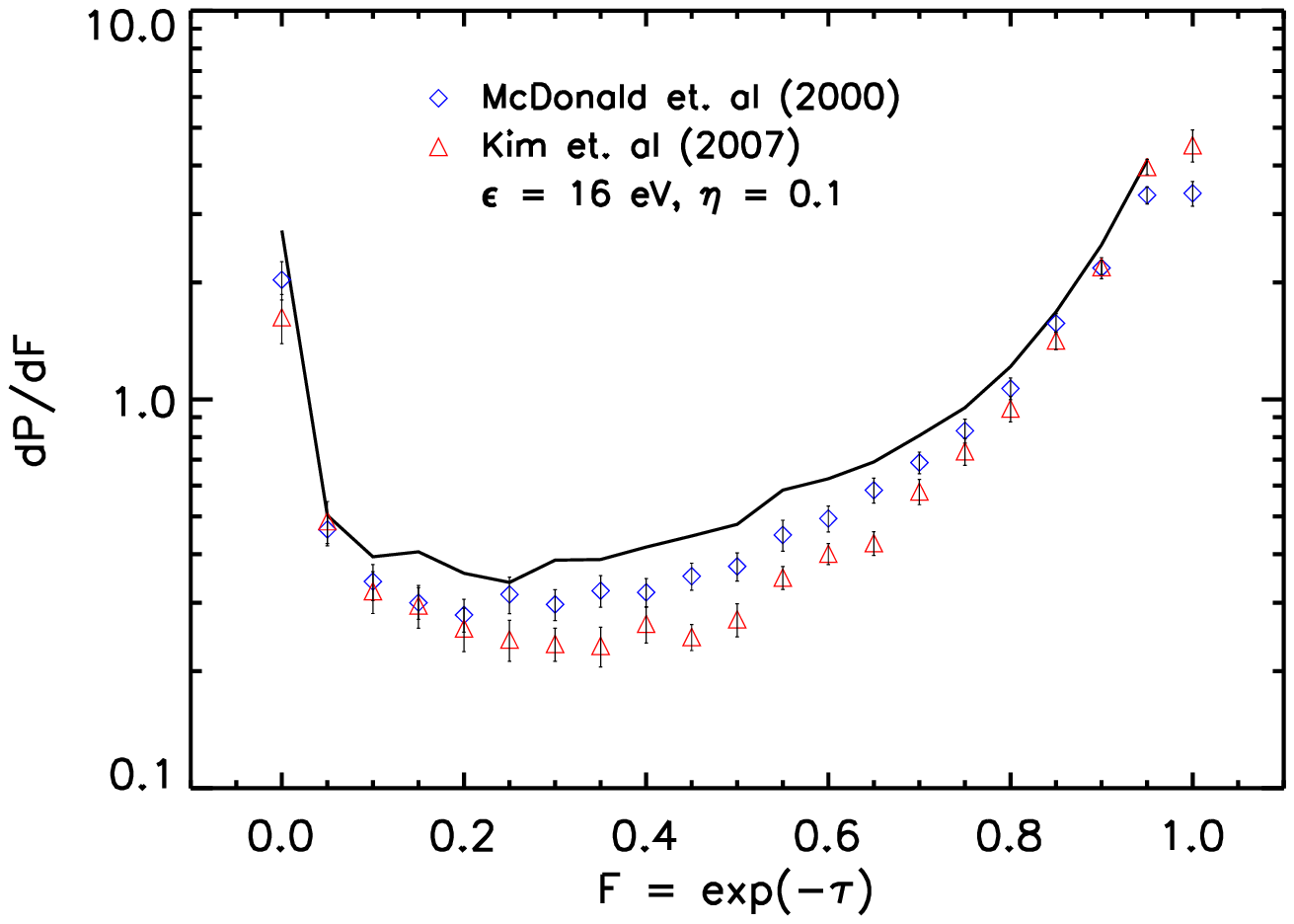}
\includegraphics[width=0.44\textwidth]{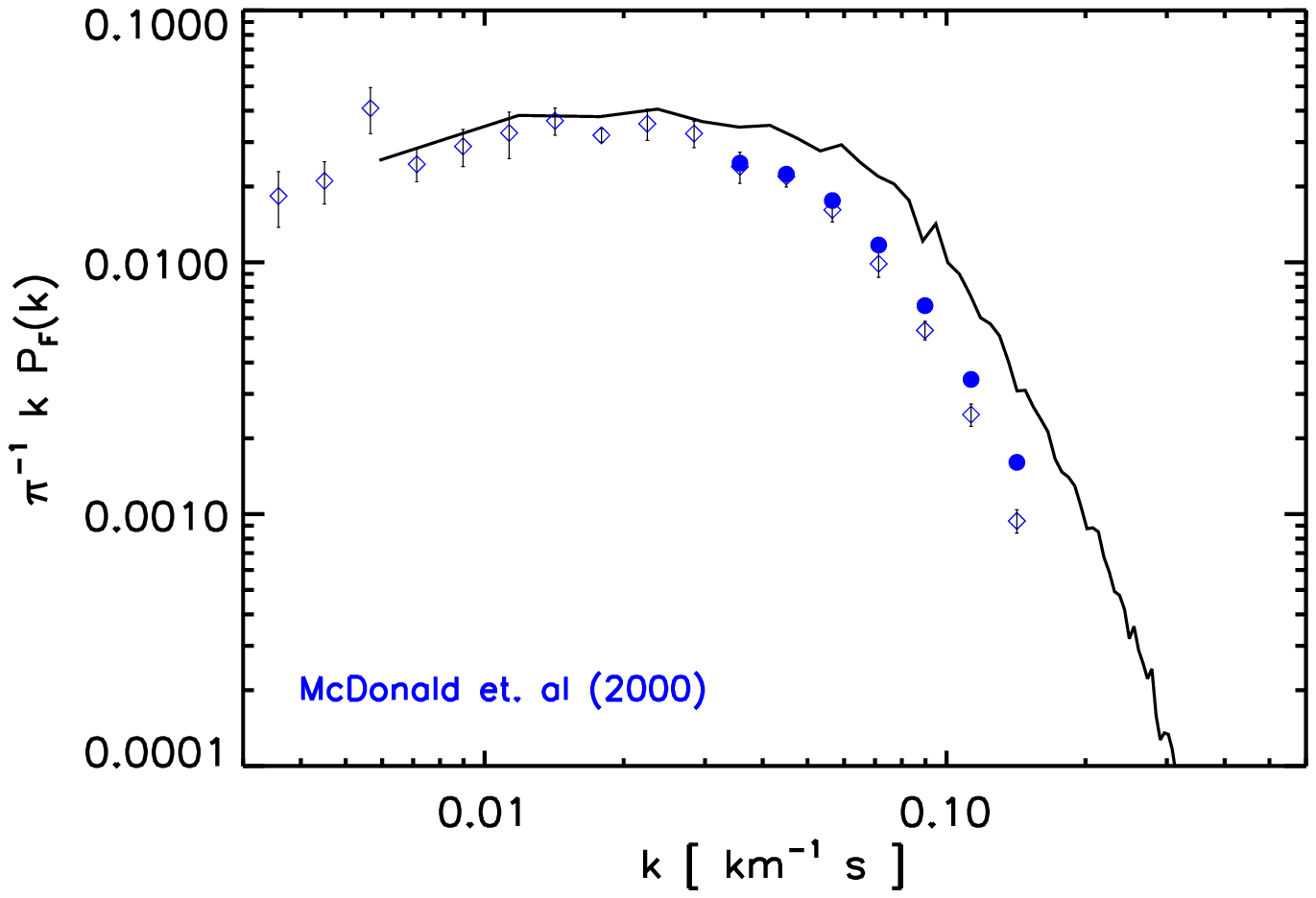}\\
\includegraphics[width=0.44\textwidth]{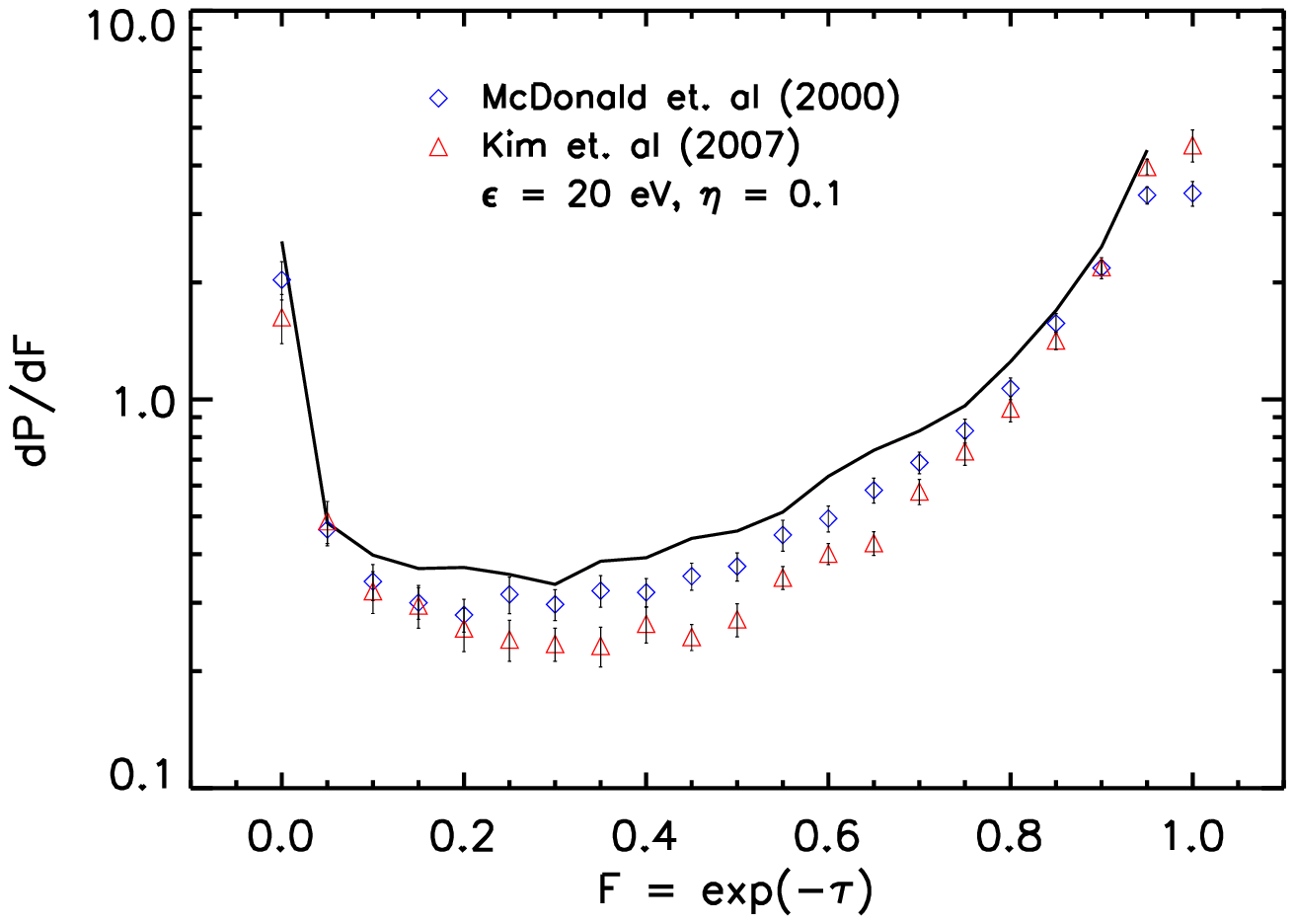}
\includegraphics[width=0.44\textwidth]{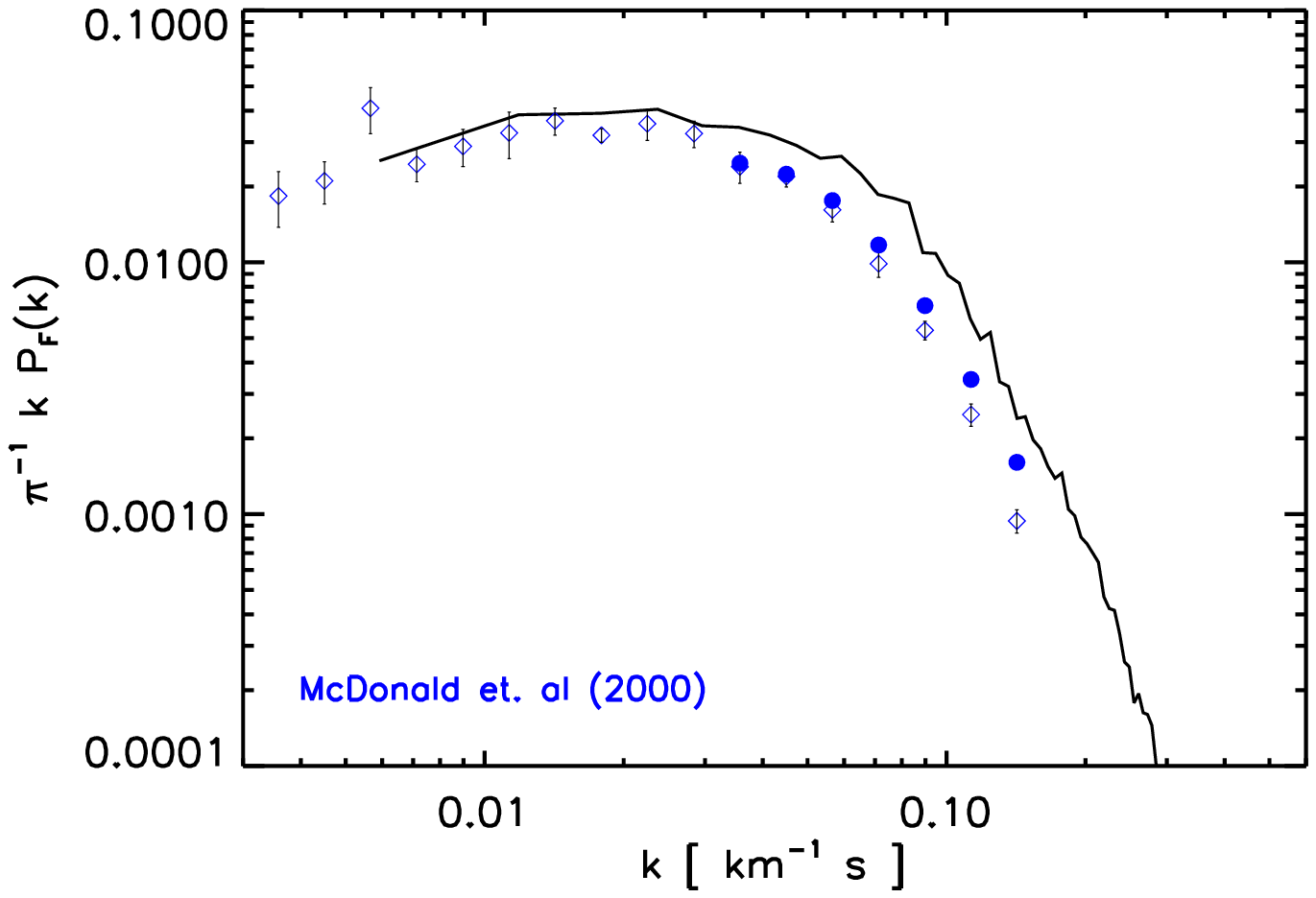}\\
\includegraphics[width=0.44\textwidth]{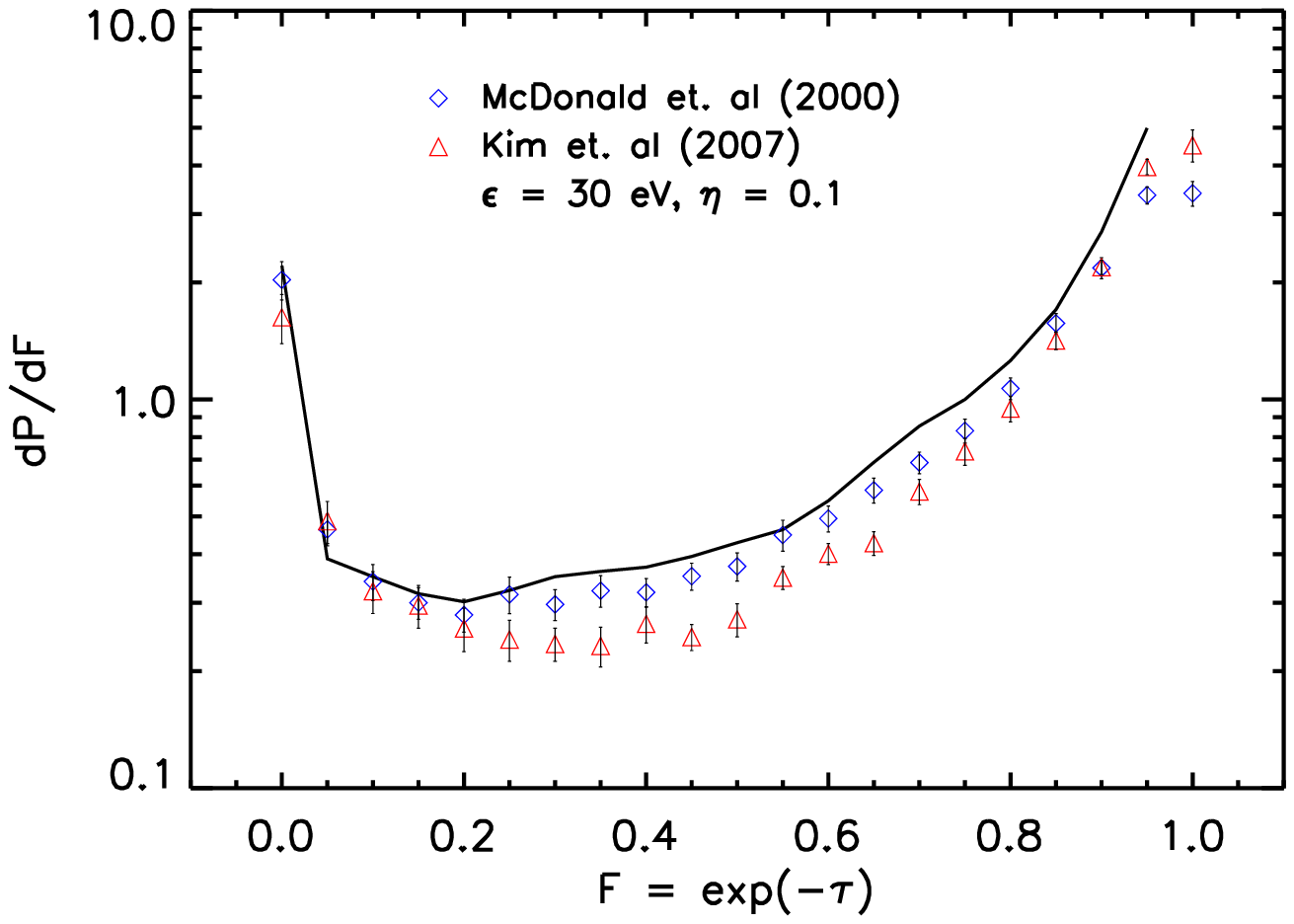}
\includegraphics[width=0.44\textwidth]{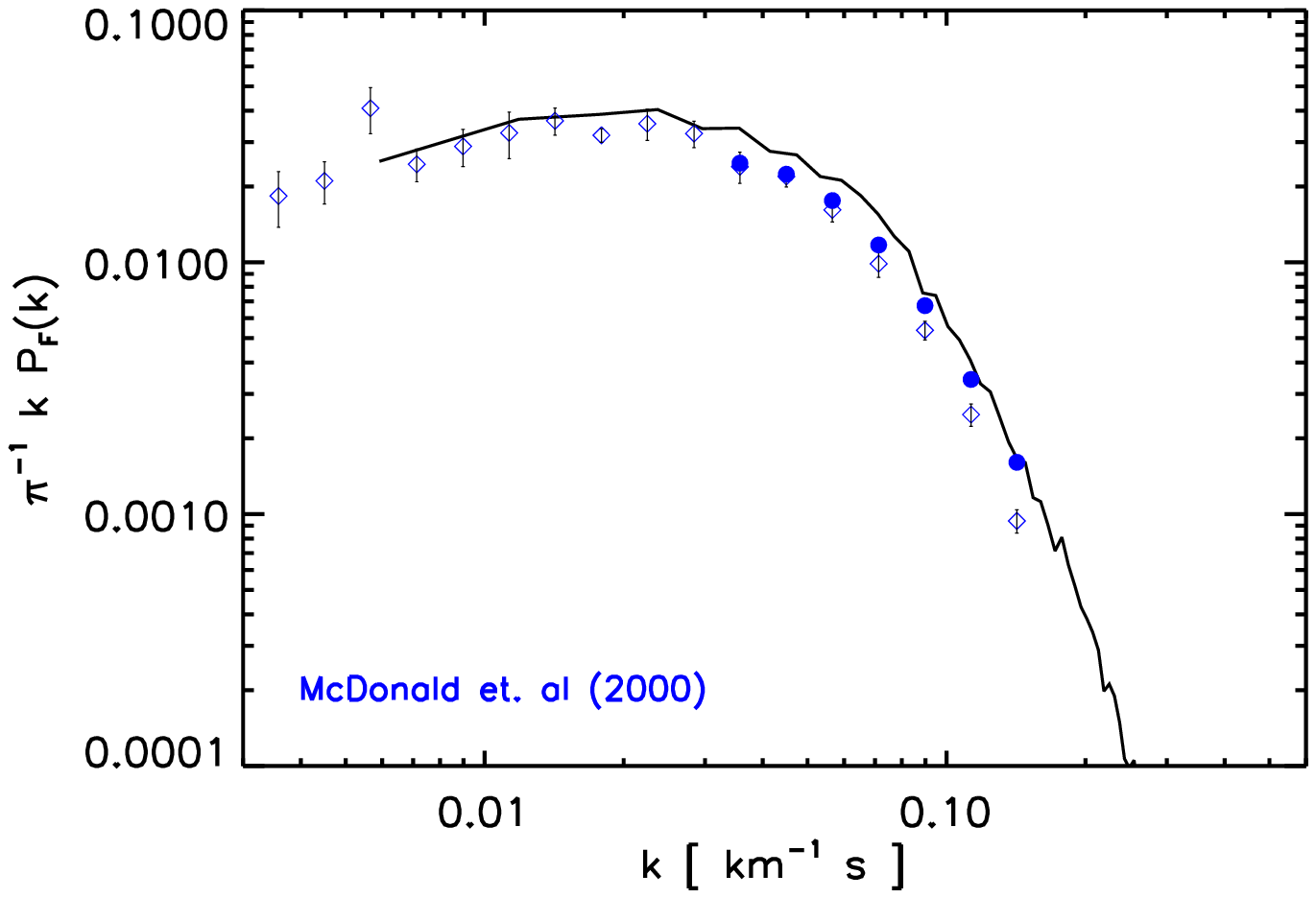}
\caption{Lyman-$\alpha$ flux probability (left) and power spectrum
  (right) for the low resolution simulation with efficiency $\eta = 0.1$, for four
  different values of the averaged excess photon energy $\tilde \epsilon
  = 6.4 \rm \,eV,\, 16\,eV,\, 20\,eV,\, 30\,eV$. The simulated data is
  compared to the observational
  result from \citet{McDonald2000} and \citet{Kim2007}. \label{fig:flux_010}} \efigs

\subsection{Simulation set}

All our simulations assume a $\Lambda$CDM universe with cosmological
parameters $\Omega_{0} = 0.27$, $\Omega_{\Lambda} = 0.73$, $\Omega_b =
0.044$, $h = 0.7$ and $\sigma_8 = 0.9$. In order to have sufficiently
high mass resolution, we follow a comparatively small region in a
periodic box of comoving size $L_{\rm box} = 10 h^{-1} \,\rm Mpc$ on a
side. This region is still sufficiently large to give a representative
account of the Lyman-$\alpha$ forest at redshift $z=3$, which is the
final time of our runs.

Our primary simulation set has $2 \times 128^3$ dark matter and gas
particles, giving a mass resolution of $3.04\times 10^7\,h^{-1}{\rm
  M}_\odot$ and $5.29\times 10^6\,h^{-1}{\rm M}_\odot$ in dark matter
and gas, respectively. A selected model was also carried out at
the much higher resolution of $2 \times 256^3$, giving a gas mass
resolution of $6.62\times 10^5\,h^{-1}{\rm M}_\odot$. The gravitational
softening was chosen as $1/35$ of the mean particle spacing in each
case, corresponding to $\epsilon=2.23\,h^{-1}{\rm kpc}$ and
$\epsilon=1.12\,h^{-1}{\rm kpc}$ in the two resolution sets. In order to save
computational time, most runs were restarted from $z=20$ with
different radiative transfer treatments, such that the higher redshift
evolution did not have to be repeated. We have
also systematically varied the `escape fraction' $\eta$ and the heating
efficiency $\tilde \epsilon$, considering the values $\eta = 0.1,\,
0.2,\, 0.3,\, 0.5,\, 1.0$ and $\tilde \epsilon = 6.4\, \rm eV,\, 16\, eV,\,
20\, eV,\, 30\, eV$ for the low resolution set. For the high
resolution run we chose the parameters $\eta=0.2$ and $\tilde
\epsilon=30 \, \rm eV$. In this way we could systematically determine the
settings that give the most promising agreement with the observations.

\section{Results} \label{sec:results}

\subsection{Hydrogen reionisation history \label{sec:history}}

When star formation starts at around redshift $z=20$ in our simulations,
the process of reionisation begins through photo-ionisation of the gas
around the star-formation sites. However, as the dense gas has a high
recombination rate, the progress in the reionisation is sensitively
determined by a competition between the luminosity of the sources, the
rate with which they turn on, and the density of the neutral gas they
are embedded in.  Using our simulation set, we first investigate the
global reionisation history and its dependence on the source efficiency
parameter $\eta$.  

Figure~\ref{fig:nHI} shows the reionisation histories of our low
resolution simulated box for several choices of $\eta$.  As expected,
the results for the reionisation redshift depend strongly on $\eta$.
For larger efficiencies, the Universe gets reionised earlier, since more
photons become available to ionise the hydrogen. This effect always
overwhelms the reduction in star formation and hence source luminosity
that an increased efficiency parameter $\eta$ also induces.  In all
cases the last phase of the Epoch of Reionisation (EoR) is rather short,
i.e.~the final 10\% of the volume transition very rapidly from being
neutral to being essentially fully ionised. The redshift of reionisation
when this occurs varies between $z\sim 5$ and $z\sim 8$ for $\eta=0.1$
and $\eta =1.0$, respectively.  In contrast, the EoR starts in general
at much higher redshift. For example, the highest efficiency $\eta =1.0$
model has already ionised nearly 30\% of the volume by redshift $z=10$.
Interestingly, there is a systematic variation of the time it takes to
complete the last phase of the EoR. This is much more rapid for the high
efficiency model than for the lower efficiency ones. Since
typical bubble sizes at redshift $z=6$ are up to $10 \, h^{-1}{\rm
  Mpc}$, we note that the size of our simulation box is really too
small to draw any definitive conclusions about the global duration of
reionisation in the Universe. A number of authors
\citep[][]{Ciardi2003,Furlanetto2005,Furlanetto2009b} have pointed out
that the local reionisation history depends on the environment, so
that, for example, the evolution of the neutral fraction in a field or
void region is different than in a proto-cluster region. As a result,
only very large simulation volumes of $100 \, h^{-1}{\rm Mpc}$ or more
on a side can be expected to yield a truly representative account of
cosmic reionization. Since at the same time an equal or better mass
resolution as we use here is required, such calculations are very
expensive and beyond the scope of this work, but the should become
feasible in the future on the next generation of high-performance
computers.

In Figure~\ref{fig:j21} we compare the ionising background for the same
simulations, which gives further interesting clues about the
reionisation histories of the different models.  The background is
computed as the volume-averaged intensity in the simulation box. 
The ionising background
is compared with an analytical estimate by \citet{Haardt1996} for the
meta-galactic ionising flux from quasars rather than stellar sources,
which is an interesting comparison point for the expected value of the
background. Clearly, lowering the efficiency $\eta$ decreases the
ionising background as well, consistent with the findings
above. Interestingly, the rapid rise of the mean background intensity
ends when reionisation is complete. From this point on, the background
shows only a weak residual evolution.

\bfig
\begin{center}
\includegraphics[width=0.47\textwidth]{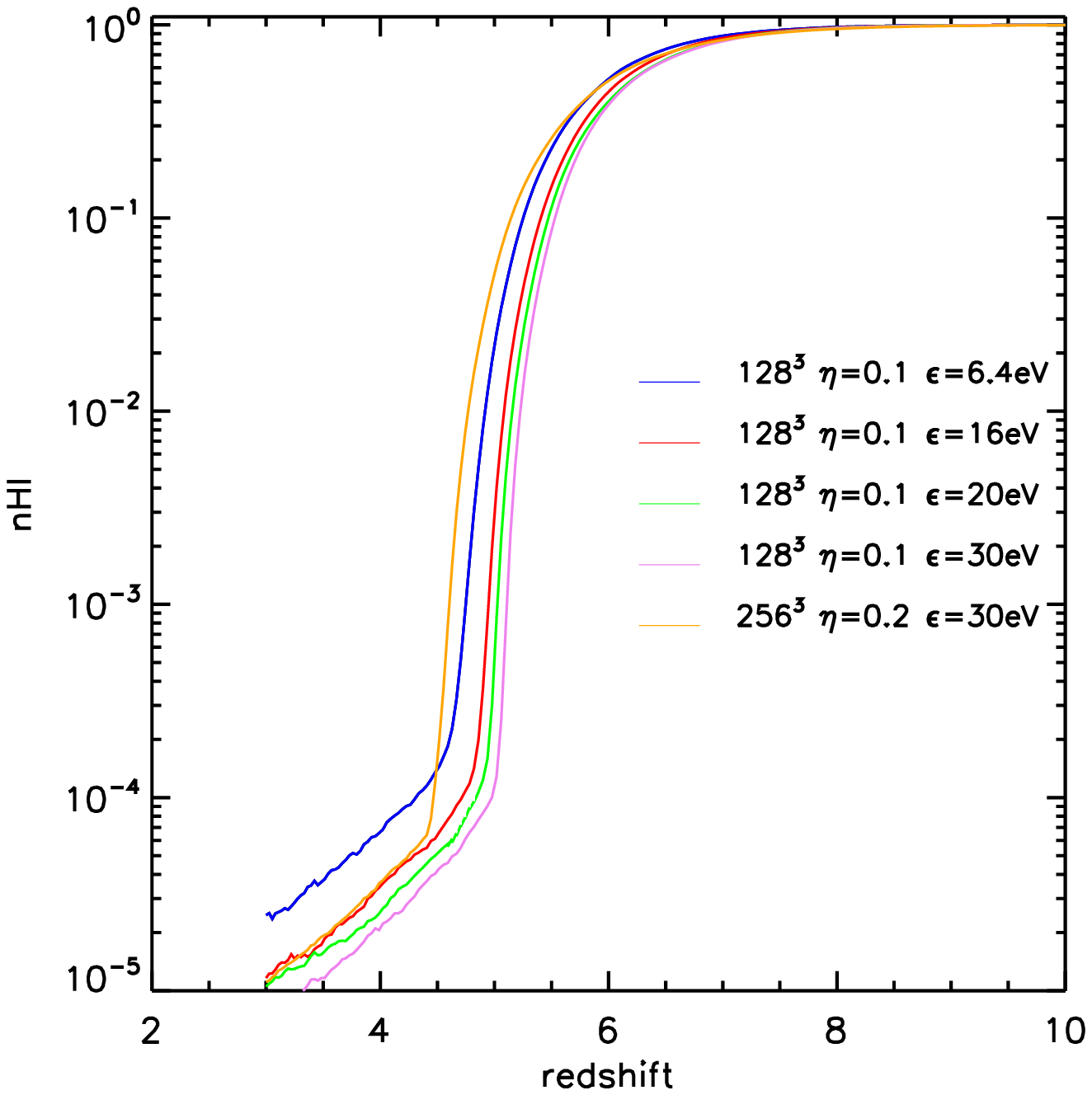}
\end{center}
\caption{Volume averaged neutral fraction as a function of redshift.
  The set of low resolution simulations with efficiency $\eta=0.1$ and
  average photon excess energy $\tilde \epsilon = 6.4\, \rm eV,\, 16\,
  eV,\, 20\, eV,\, 30\, eV$ is compared with the high resolution run
  with $\eta = 0.2 \rm$ and $\tilde \epsilon=30 \, \rm eV$. When a
  higher energy per ionising event is injected, the universe gets
  ionised slightly earlier since higher temperatures help to maintain
  higher ionised fractions.\label{fig:nHI_esc010}} \efig

The time evolution of the temperature, ionised fraction and density
fields in a representative simulation model ($\eta=0.1$ and $\tilde
\epsilon=30 \, \rm eV$) is illustrated in
Figure~\ref{fig:dens_evolve}. The different panels correspond to slices
through the middle of the simulation box, between redshifts $z=7.2$ and
$z=4$, from the top row to the bottom row.  It is seen that the ionised
regions start to grow first around high density peaks, where the star
forming regions are concentrated. Then the radiation diffuses into the
inter-cluster medium. The filaments remain less ionised than the voids
for a while, since their density is much higher. Initially the photons
heat up the gas in the ionised region to temperatures slightly above
$10^4 \rm K$. As the ionising background declines due to the expansion
of the Universe and the drop of star formation, the heating becomes
less effective and the temperature of the highly ionised gas in the
voids drops somewhat as a result of the expansion cooling.

\bfig
\begin{center}
\includegraphics[width=0.47\textwidth]{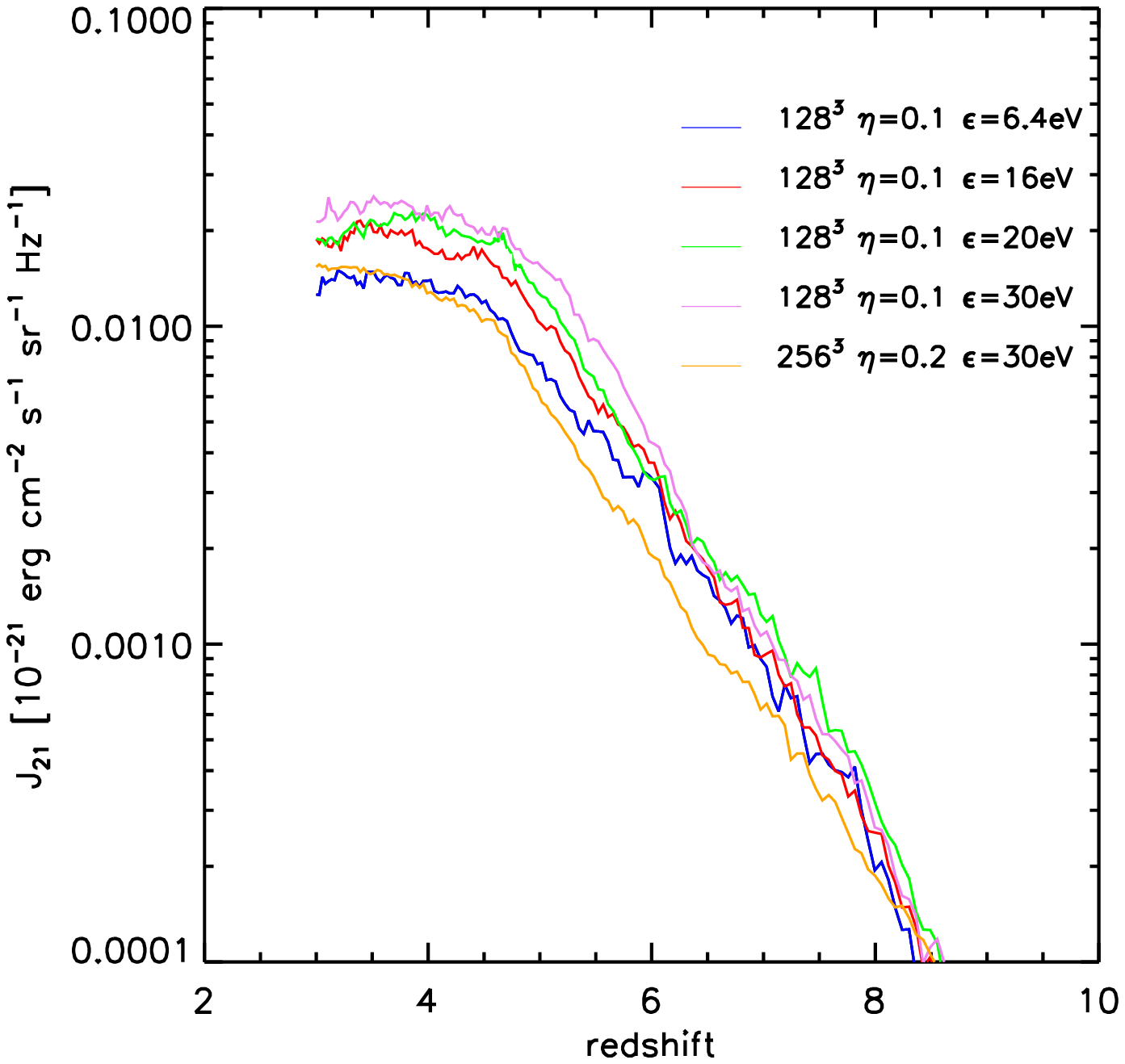}
\end{center}
\caption{Ionising background as a function of redshift. The set of low
  resolution simulations with efficiency $\eta=0.1$ and
  averaged photon excess energy $\tilde \epsilon = 6.4\, \rm eV,\, 16\,
  eV,\, 20\, eV,\, 30\,eV$ is compared with the high
  resolution run with $\eta = 0.2 \rm$ and $\tilde \epsilon=30 \, \rm eV$. 
  For a higher injected energy per
  ionisation event, the background also increases as a result of the
  higher temperature, which leaves more photons unabsorbed so that they
  can contribute to a higher level of the ionising
  background. \label{fig:j21_esc010}} 
\efig

In Figure~\ref{fig:q_rho}, we present a contour plot of the neutral
fraction versus the over-density of the gas in the representative
model for two different redshifts,
at $z = 6.4$ before reionisation is completed, and at $z=3$ after
reionisation is completed. There is a clear dependence of the neutral
fraction on over-density in both cases. The high density regions
around star-forming matter are ionised very quickly. The average
density regions, e.g. filaments, tend
to be less ionised and get ionised after the low density
regions. However, note that in the star-forming tale of the diagram all
the gas is ionised. This is here due to the star formation scheme
adopted in {\small GADGET}, where star-forming gas particles are
assigned a mean mass-weighted temperature which is so high that all this
gas is formally collisionally ionised. We note that these results are
very similar to the ones reported by \citet{Gnedin2000}, where a similar
relation between neutral fraction and density is found. However, we are
able to probe somewhat higher densities thanks to better spatial
resolution of our simulations.

\subsection{Lyman-$\alpha$ forest \label{sec:Lyman-a}}

We next turn to an analysis of the thermal state of the intergalactic
medium left behind at $z=3$ by our self-consistent reionisation
simulations. To this end we use the gas density, gas temperature, gas
velocity and ionisation state of the gas in the simulation box and
compute Lyman-$\alpha$ absorption spectra for random lines of sight.  We
then compare the statistics of these artificial absorption
spectra with observational data on the Lyman-$\alpha$ forest at this
epoch, as given by \citet{McDonald2000} and \citet{Kim2007}. Here we assume that the main
source of ionising sources is stellar and thus discard any
contribution from a quasar-type spectra, in agreement with
\citet{Madau1999}. We also note that some discrepancies are possible
due to the fact that helium is not photo-ionised in our simulations, but
only collisionally ionised.

Our simulations have the necessary gas mass resolution at
redshift $z=3$ required to reproduce realistic Lyman-$\alpha$
absorption in the low density regions \citep{Bolton2009}. However, we
can not match the required simulation volume of size $\sim 40 \,
h^{-1} {\rm Mpc}$ to properly sample the largest voids. This can have
a significant effect on our predictions of the Lyman-$\alpha$ flux
probability distribution and power spectra.

Figure~\ref{fig:flux_020} shows the flux probability distribution
function (PDF) and flux power spectrum for our simulation model, where an
efficiency parameter of $\eta = 0.2$ and averaged photon excess
energy $\tilde \epsilon = 6.4 \, \rm eV$ were adopted. For this choice, we
achieve the best fit to the flux PDF. However, for all the other models
the power spectrum is over-predicted at high wave numbers. We suggest
that this overestimation of the power spectrum is due to the
insufficient heating of the gas in low density regions, causing an
excess of small-scale structure in the Lyman-$\alpha$ forest.

\bfig
\begin{center}
\includegraphics[width=0.47\textwidth]{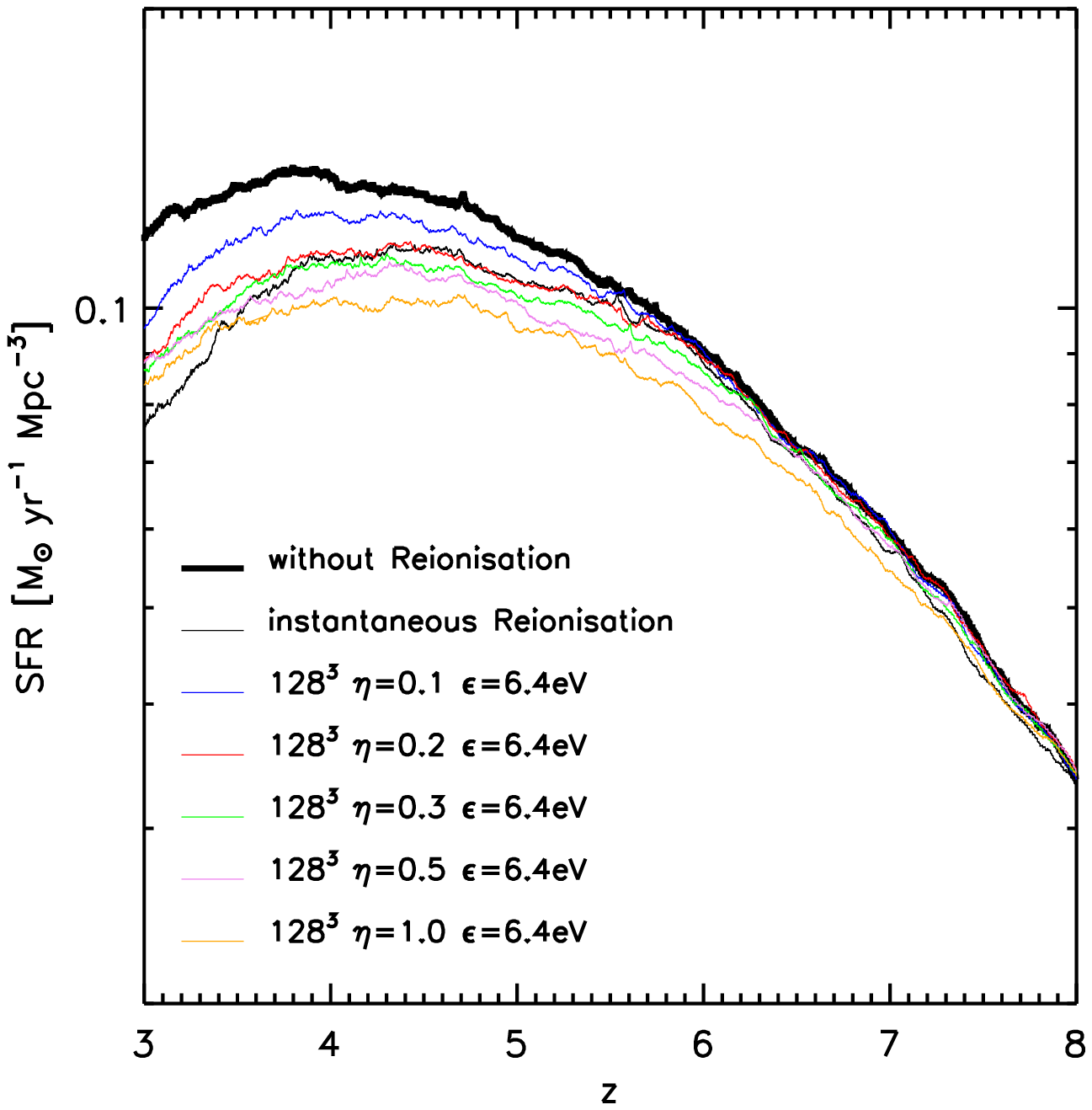}
\end{center}
\caption{Star formation rate density as a function of redshift for
  the low resolution simulation set at different efficiencies of $\eta =
  0.1,\, 0.2,\, 0.3,\, 0.5,\, 1.0$. 
  The results are compared to the SFR history of a low resolution simulation
  with instantaneous reionisation at $z=6$ and photo-heating by a
  \citet{Haardt1996} ionising background (thin black line), and a
  simulation with neither reionisation nor photo-heating (thick black line). The
  photo-heating from stellar sources decreases star formation, as
  suggested by \citet{Pawlik2009}. As the escape efficiency gets higher,
  this effect becomes progressively stronger. \label{fig:SFR}} 
\efig

In order to examine this effect further, we vary the photon
excess energy $\tilde \epsilon$ used in the photo-heating and examine
the influence this has on the flux probability and power
spectrum. There are two possible reasons why our simulations
underestimate the photo heating. First, we expect that some
non-equilibrium effects in the photo-heating are treated inaccurately
due to our implicit treatment of the radiation transport and chemistry
\citep[e.g.][]{Bolton2009}. Second, photo-heating is different in
optically thin and optically thick regions. For example, in an
optically thick region the average photon excess energy obtained from
Eqn.~(\ref{eqn:epsilon}) is $\tilde \epsilon = 29.9 \, \rm eV$. It is
however likely that our approximative radiative transfer scheme leads
to inaccurcies in the effective heating rates of regions of different
optical depths, due to the varying accuracy of the scheme in different
regimes. Part of these inaccuracies can be absorbed into a suitably
modified value of the effective heating rate $\tilde \epsilon$. To
explore the full range of plausible values, we therefore vary the
values for $\tilde \epsilon$ as follows: $\tilde \epsilon = 6.4 \,\rm
eV, \, 16 \, eV, \, 20\, eV$ and $30\, {\rm eV}$. We aim to bracket
what can be expected when non-equilibrium effects are fully taken into
account in future treatments, and want to identify the case that
provides the best representation of the Universe at redshift $z=3$.
  
Figure~\ref{fig:flux_020} shows the flux PDF
and power spectra for these different heating values. Clearly, the high
wave number region of the flux power spectrum is strongly influenced by
the amount of injected heat energy into the gas, and the increase of the
temperature also affects the flux probability distribution.  For the low
efficiency of $\eta = 0.1$, there is a substantial mismatch already in
the flux PDF, simply because there is too little ionisation overall so
that the mean transmission ends up being too low.  However, as the
adopted photo-heating energy increases, the gas is getting hotter and is
able to stay ionised longer due to the higher temperatures, yielding a
better fit to the flux PDF. At the same time, small-scale structure in
the flux power spectrum is erased due to thermal broadening, bringing
the simulations into agreement with the observation. This shows the
power of detailed Lyman-$\alpha$ data to constrain simulations of the
reionisation process. In our current models we need to adopt a quite
extreme heating efficiency of $30\,\rm eV$ combined with a low `escape
fraction' of $\eta = 0.1$ to achieve a good match to the data.

\bfig
\begin{center}
\includegraphics[width=0.47\textwidth]{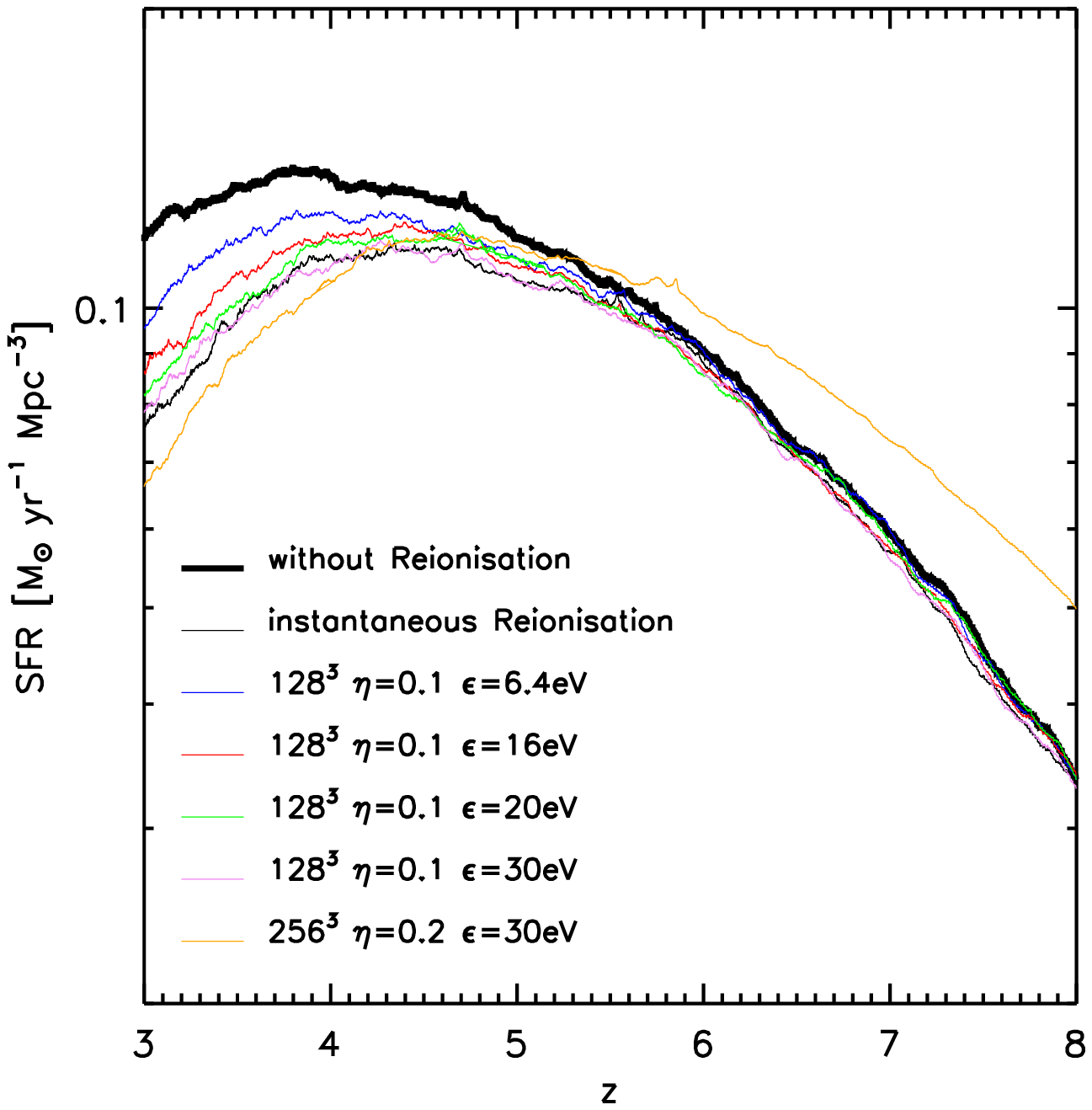}
\end{center}
\caption{Star formation rate density as a function of redshift. The
  set of low resolution simulations with efficiency $\eta=0.1$ and
  averaged photon excess energy
  $\tilde \epsilon = 6.4 \, \rm eV,\, 16\, eV,\, 20\, eV, \, 30\,eV$
  are compared to the high resolution run with $\eta = 0.2$ and
  $\tilde \epsilon=30 \, \rm eV$. The results are compared to the SFR history
  of a low resolution simulation
  with instantaneous reionisation at $z=6$ and photo-heating by a
  \citet{Haardt1996} ionising background (thin black line), and a
  simulation with neither reionisation nor photo-heating (thick black
  line). The star formation decreases with increasing heating energy,
  as expected. For the low resolution run with $30\,\rm eV$, the result of the
  self-consistent radiative transfer calculation matches the simulation
  with instantaneous reionisation. The high resolution simulation SFR
  is higher at higher redshift due to better resolution and agrees
  well with the other results at redshifts less than $z=6$.\label{fig:SFR_esc010}}
\efig

\bfigs
\includegraphics[width=0.44\textwidth]{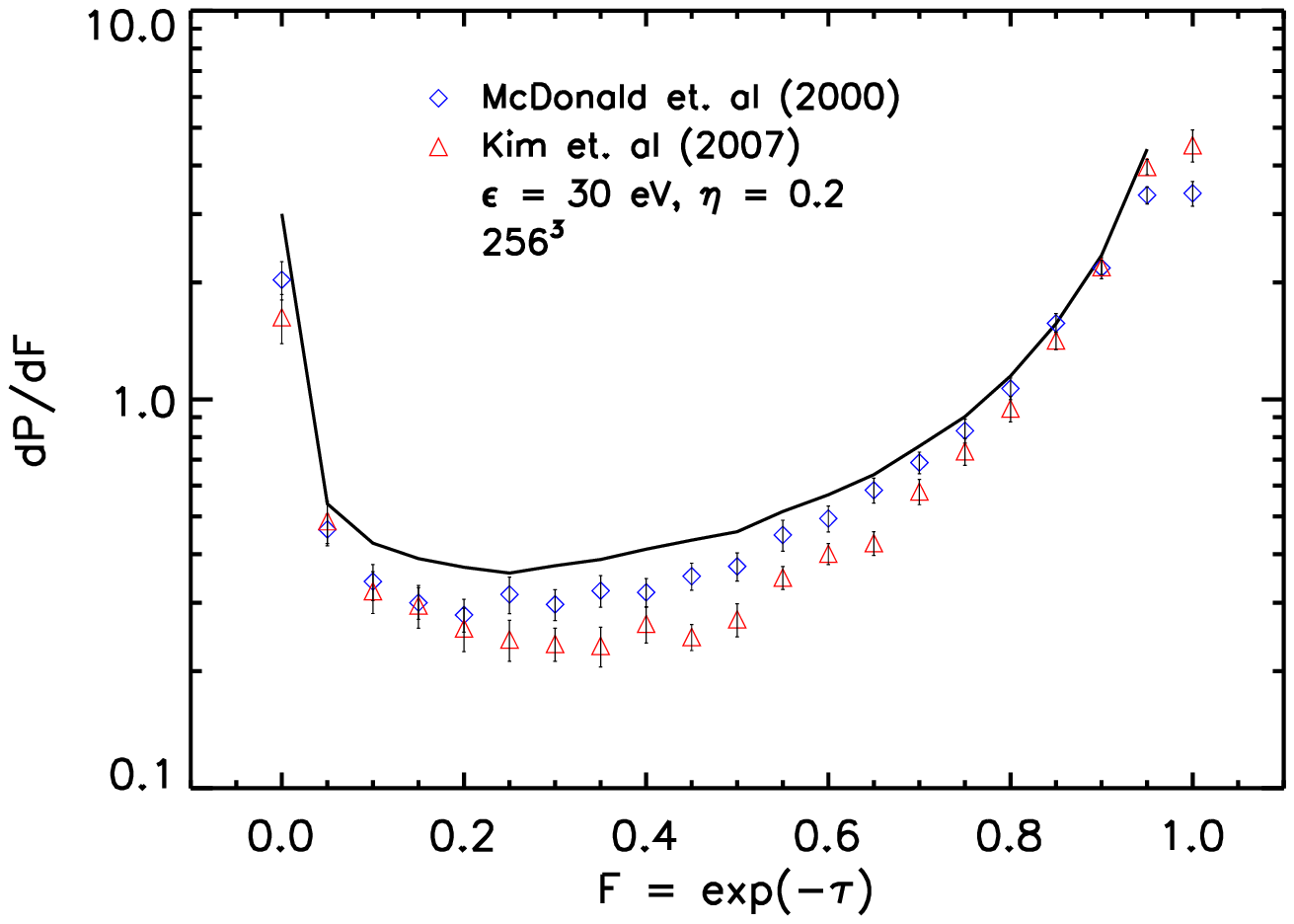}
\includegraphics[width=0.44\textwidth]{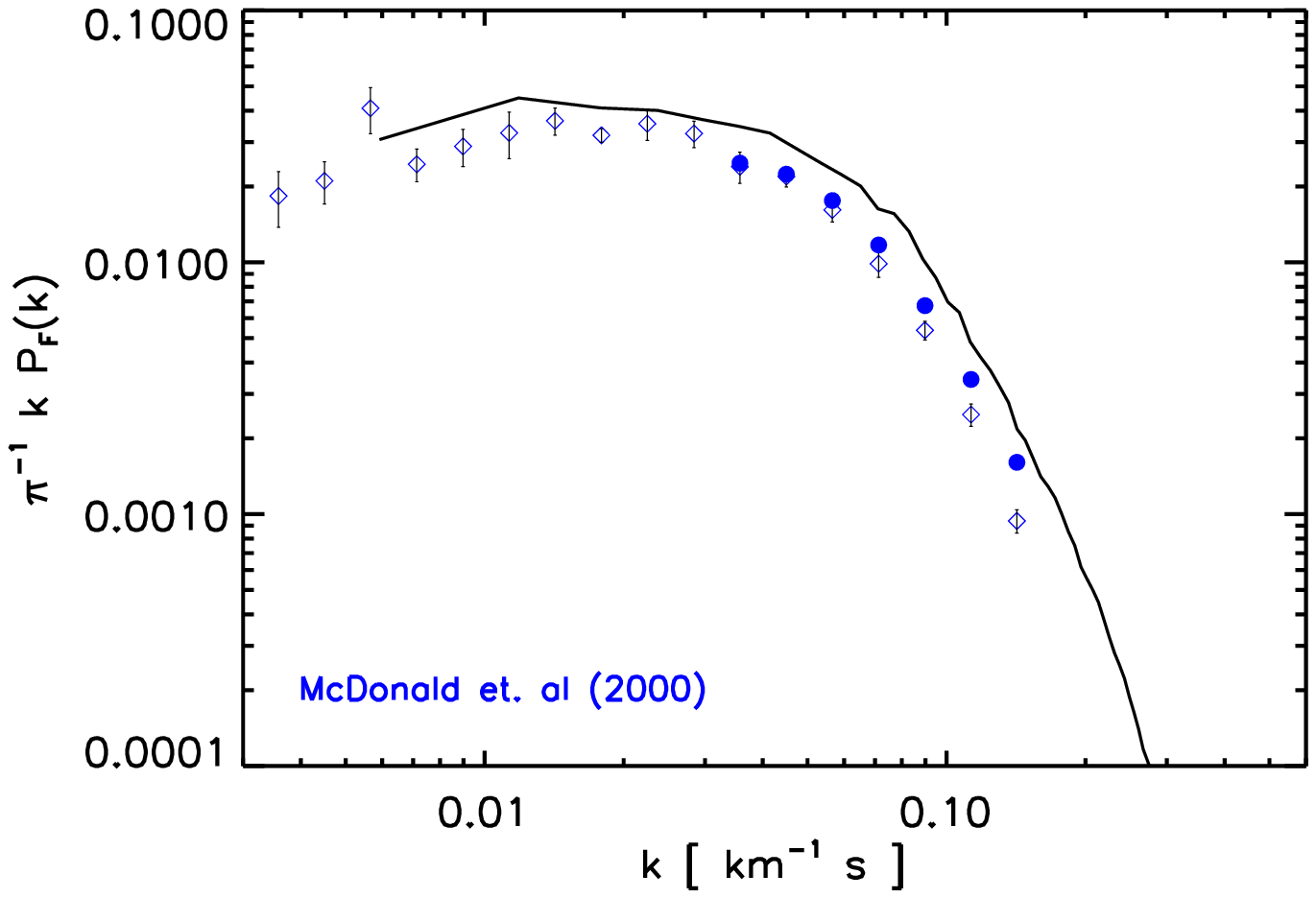}
\caption{Lyman-$\alpha$ flux probability (left) and power spectrum
  (right) for the high resolution simulation with efficiency $\eta =
  0.2$ and averaged excess photon energy $\tilde \epsilon = 30 \, \rm
  eV$, compared to observational results from
  \citet{McDonald2000} and \citet{Kim2007}. \label{fig:lyman256}} \efigs

In Figures~\ref{fig:nHI_esc010} and \ref{fig:j21_esc010} we compare the
impact of the different photo-heating efficiencies on the evolution of
the neutral volume fraction and the ionising background. We also show
for comparison the results from our high resolution simulation, which
is discussed below in the text. As expected,
an increase in the heating energy leads to a slightly earlier
reionisation and to a slightly elevated ionising background flux. Both
of these effects can be readily understood from the higher gas
temperature produced in the ionised gas when the higher heating
efficiency is adopted. However, the effect is quite weak, and very much
smaller than the changes resulting from a different choice of $\eta$.

We have also measured the Thomson electron scattering optical depth in
our high resolution simulation and found it to be $\tau_{\rm es} =
0.049$, which is smaller than the WMAP7 value $\tau_{\rm es}^{\rm
  WMAP} = 0.088 \pm 0.015$ \citep{wmap7}. This discrepancy is,
however, not critical since the simulated volume is too small to
obtain a realistic value and we have also not included photo-ionisation of
helium.

\subsection{Feedback from reionisation \label{sec:feedback}}

Reionisation can in principle exert a strong feedback effect on the gas
through the temperature increase induced by photo-heating.  As the gas
temperature increases, the gas densities will be lowered through
pressure effects. The gas will then cool and collapse more slowly, such
that the star formation rate is ultimately reduced. Especially small
dark matter halos should be sensitive to this effect. In the extreme
case of halos that have virial temperatures comparable to or only
slightly larger than the temperature reached by the gas through
reionisation, the UV radiation may even completely suppress atomic
cooling and efficient star formation. This effect is often invoked to
explain why so many of the dark matter satellites expected in
$\Lambda$CDM in the halos of ordinary $L_\star$ galaxies are apparently
largely devoid of stars.

In order to highlight the radiative transfer effects on the star
formation in galaxies, we compare our simulations with two fiducial
models where no radiative transfer is used.  The first is a simulation
simply without any photo-heating of the gas, while the second one is a
simulation where reionisation is induced by an externally imposed,
spatially homogeneous UV background based on a modified
\citet{Haardt1996} model that causes reionisation and an associated photo-heating of the
gas at $z=6$ \citep[for details see][]{Dave1999}. The latter model corresponds to
the standard approach applied in many previous hydrodynamical simulation
models of galaxy formation \citep[e.g.][]{Tornatore2003, Wadepuhl2010}.

In Figure~\ref{fig:SFR}, we compare our results for the cosmic star
formation rate density evolution as a function of the adopted efficiency
parameter $\eta$ (for the low resolution simulation set). 
We also include the two fiducial comparison models as
limiting cases. As we increase the escape efficiency of the ionising
radiation, the star formation drops, as expected, since this makes more
photons available to photo-heat the gas. We note that our results for
the SFR are always lower than the fiducial simulation where no
photo-heating is included at all, consistent with findings by
\citet{Pawlik2009}. Towards lower redshift, the reduction of the SFR due
to the radiation field becomes progressively larger. The run with
$\eta=0.2$ quite closely corresponds to the simulation with the imposed
reionisation epoch, but starts to slightly differ at redshifts $z < 4$.

We also carry out a corresponding comparison for a low resolution simulation set with
constant efficiency $\eta=0.1$ but different values for the photon
excess energy.  In Figure~\ref{fig:SFR_esc010} we show the results for
the SFR, again including the two fiducial models as limiting cases for
comparison. The results confirm the expectation that an increase of the
photon excess energy decreases the star formation rate density.
Interestingly, the model that best reproduced the Lyman-$\alpha$ power
spectrum observations, the one with $\tilde \epsilon = 30 \, \rm eV$, quite
closely follows the star formation rate density obtained for the
fiducial model where reionisation is imposed at $z=6$.

For our high resolution run we chose to repeat the simulation with
averaged photon excess energy $\tilde \epsilon = 30 \, \rm eV$ and
adopt a higher escape fraction $\eta = 0.2$ rather than $\eta=0.1$. In
this way we make sure we account for the trapping of photons in high
density peaks, which were not present in the low resolution runs. In
Figure~\ref{fig:SFR_esc010} we show how the star formation rate
history compares to the ones from the low resolution runs. They are in
good agreement, except for the higher redshift, where the high
resolution captures more star formation, as expected
\citep{Springel2003b}.  The Lyman-$\alpha$ forest flux probability and
power spectrum at $z=3$ for this simulation are shown in
Figure~\ref{fig:lyman256}.  While the simulated data is in reasonable
qualitative agreement with the observational results from
\citet{McDonald2000} and \citet{Kim2007}, it does not provide in this
case a detailed fit 
within the error bars, again highlighting that simultaneously
accounting for the cosmic star formation history, cosmic reionisation
and the state of the IGM at intermediate redshifts provides a powerful
constraint on self-consistent simulations of galaxy formation and
reionisation.

\bfig
\includegraphics[width=0.47\textwidth]{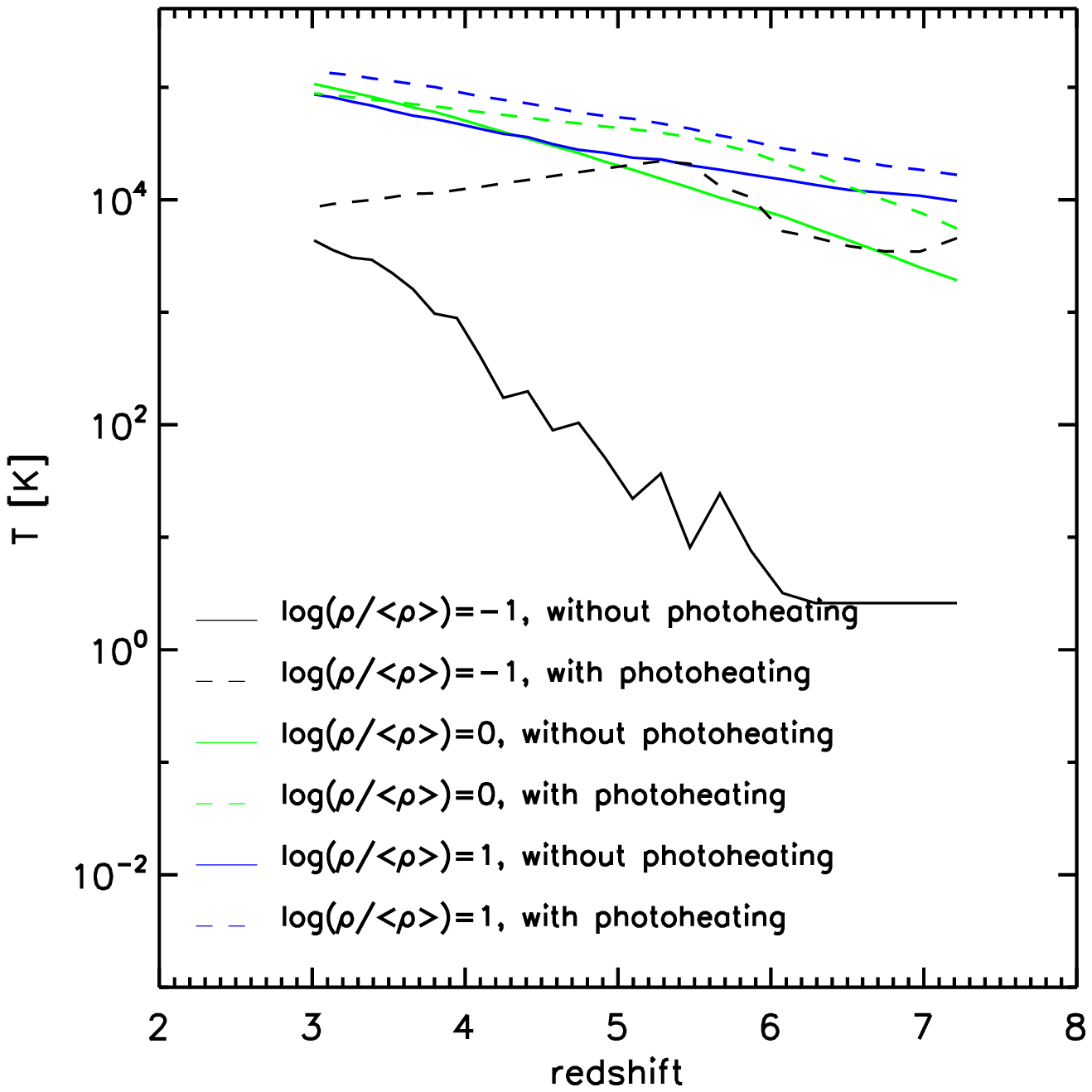}
\caption{Evolution of the mass averaged temperature at three different
  over-densities $\log_{10}(\rho/\left<\rho\right>)=-1, 0, 1$ for the
  low resolution simulation with $\eta=0.1$ and $\tilde \epsilon=30 \,
  \rm eV$, with and without photo-heating. The strongest effect is observed in
  the low density gas, which is heated by the photons much more than
  the higher density gas. At all densities, however, photo-heating
  increases the temperature, as expected.\label{fig:temperature}}
\efig

\bfig
\includegraphics[width=0.47\textwidth]{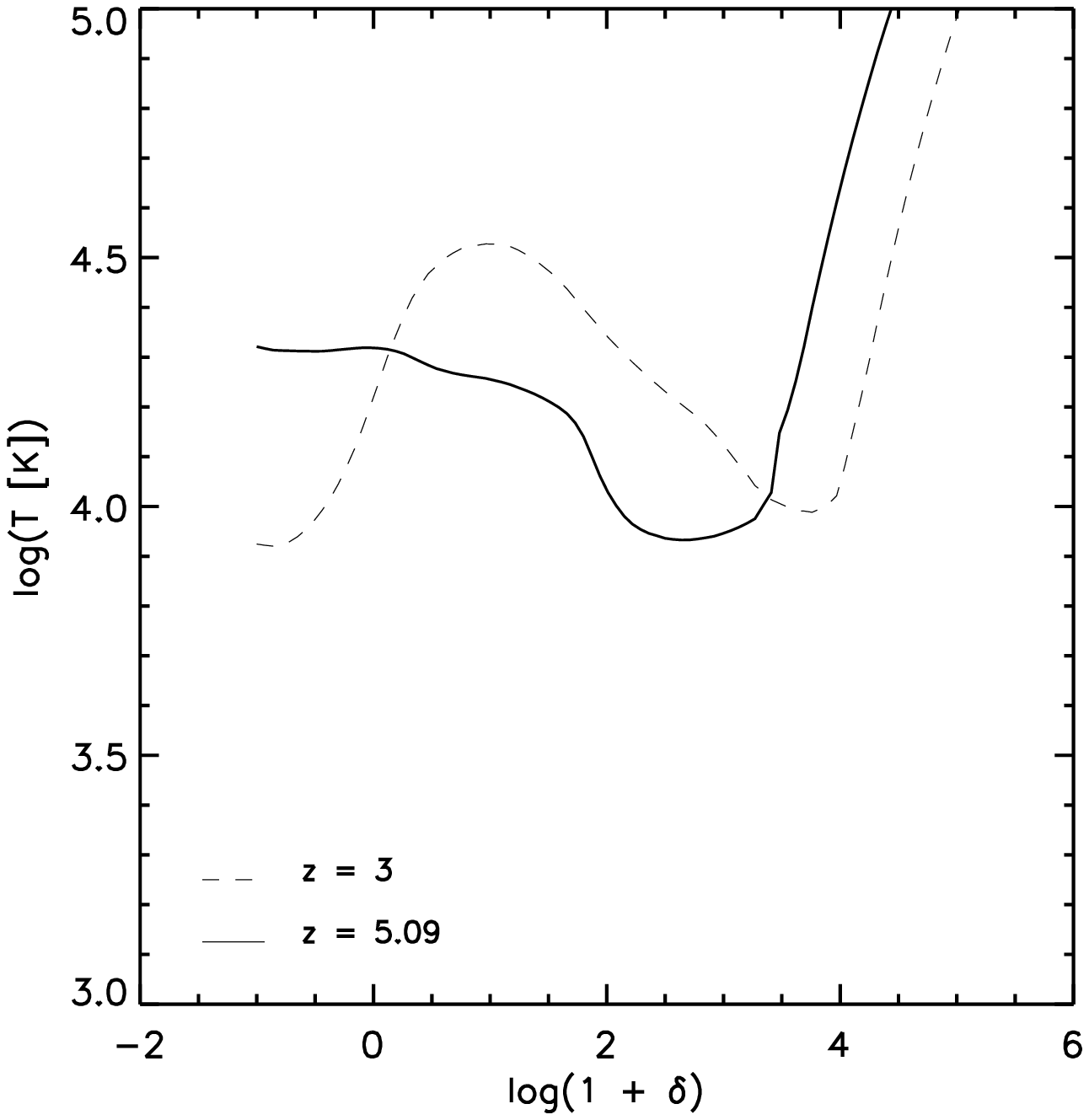}
\caption{Median temperature of the gas as a function of over-densities
  $\log_{10}(1+\delta)$ at redshifts $z=5.09$ and $z=3$ for the high
  resolution simulation with $\eta=0.2$ and $\tilde \epsilon=30 \, \rm
  eV$. At redshift $z=5.09$, shortly after reionisation is completed,
  the temperature at low densities is clearly higher than that at higher
  densities (apart from the gas in the star-forming phase). This can be
  interpreted as an inverted equation of state. At lower redshift the
  relation reverts again to normal form as the gas in the low density
  regions cools down adiabatically due to the expansion of the
  Universe. \label{fig:temp_inv}} \efig

\bfig
\includegraphics[width=0.47\textwidth]{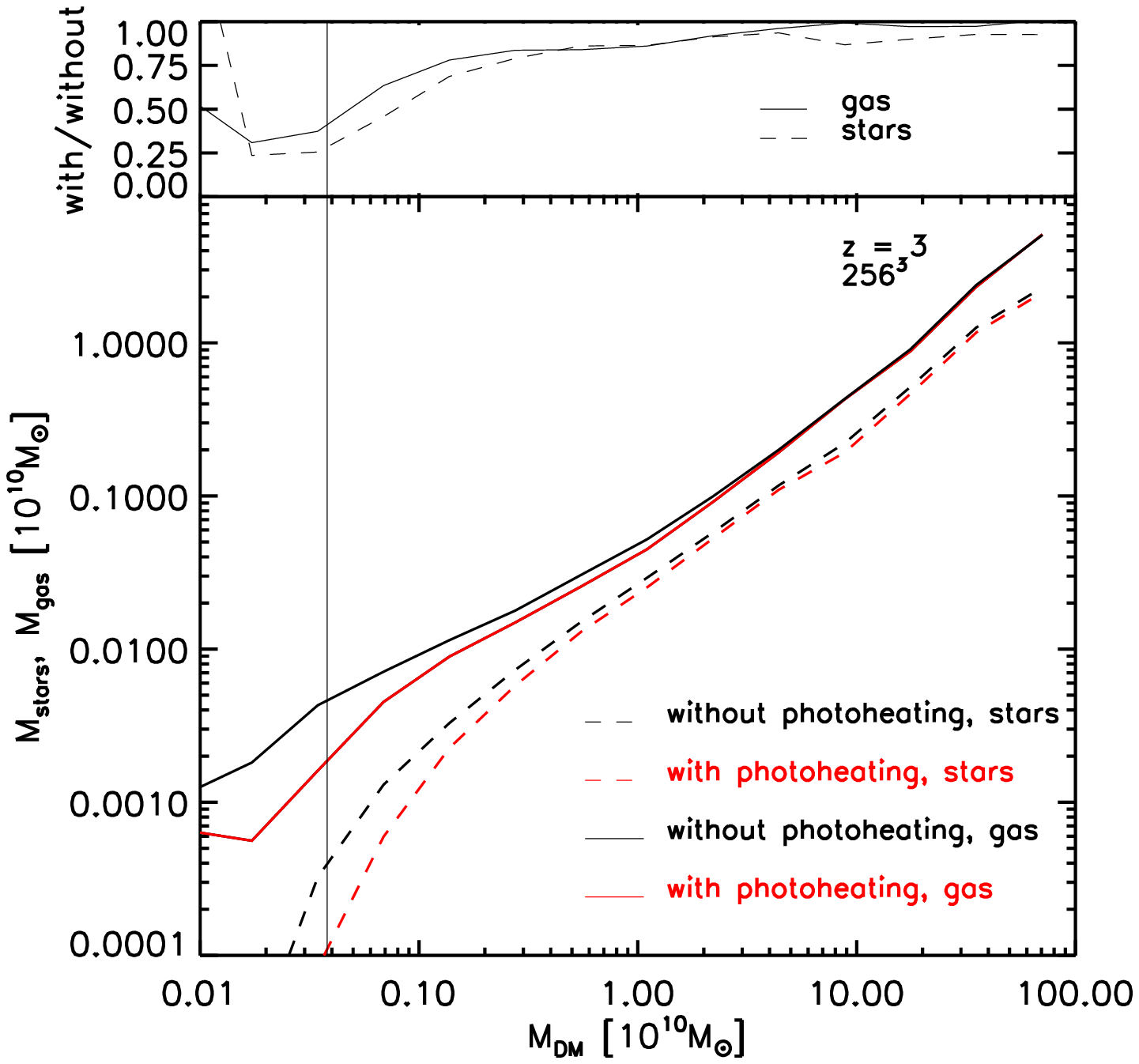}
\caption{Mean stellar and gas masses as a function of the DM halo mass
  at $z=3$ in the high resolution simulation. The black vertical
  corresponds to a mass of 100 DM particles, which can be taken as an
  (optimistic) resolution limit of the simulation.  Photo-heating
  slows down the collapse of gas in halos, which in turn also
  decreases their stellar and gas masses. The effect becomes stronger for
  low mass DM halos. \label{fig:GSvsDM}} \efig

\bfig
\includegraphics[width=0.47\textwidth]{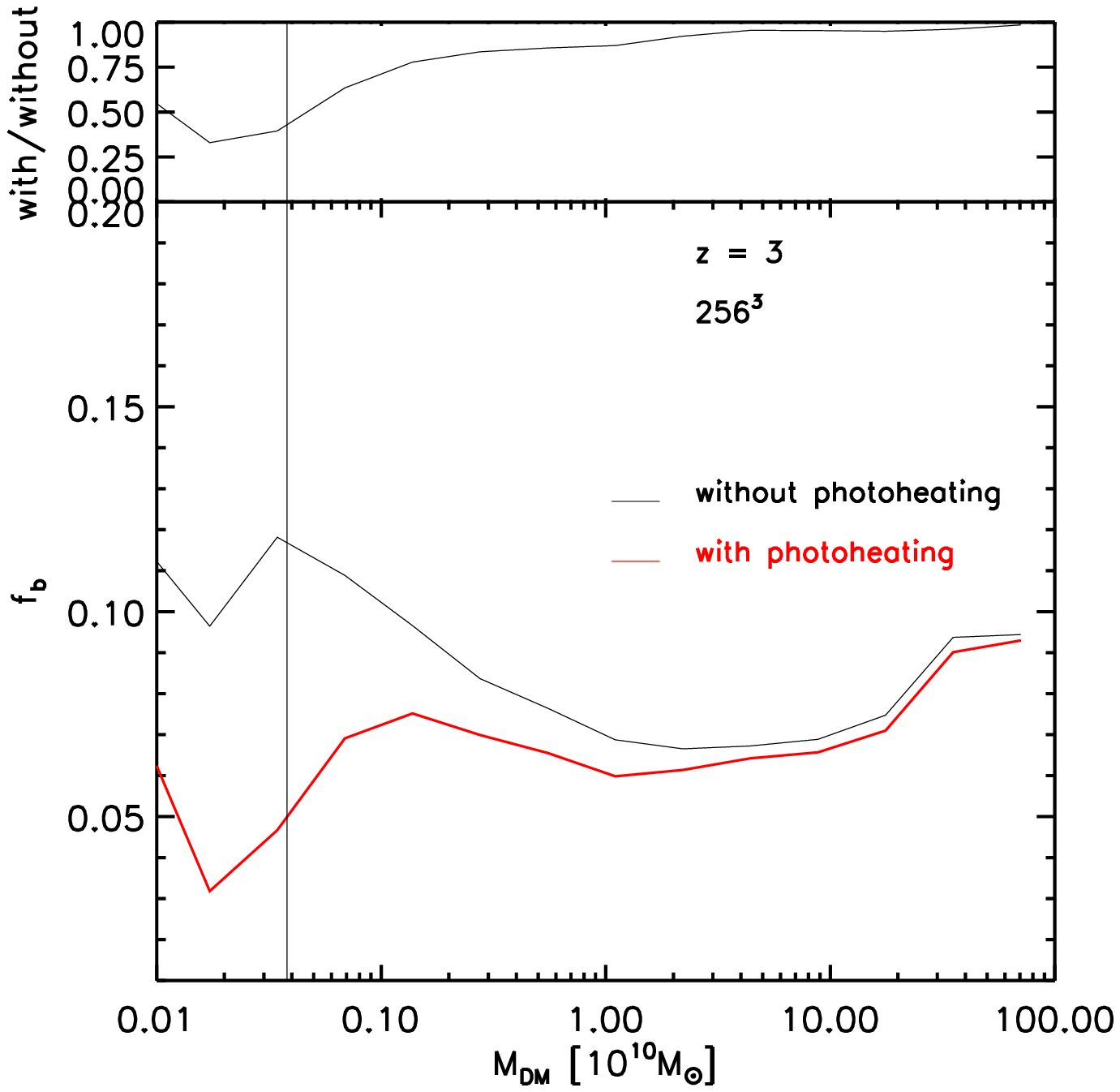}
\caption{Baryon fraction as a function of the DM halo mass at $z=3$ in
  the higher resolution simulation.  The black vertical
  corresponds to a mass of 100 DM particles.
\label{fig:f_b_256}} \efig

\bfig
\includegraphics[width=0.47\textwidth]{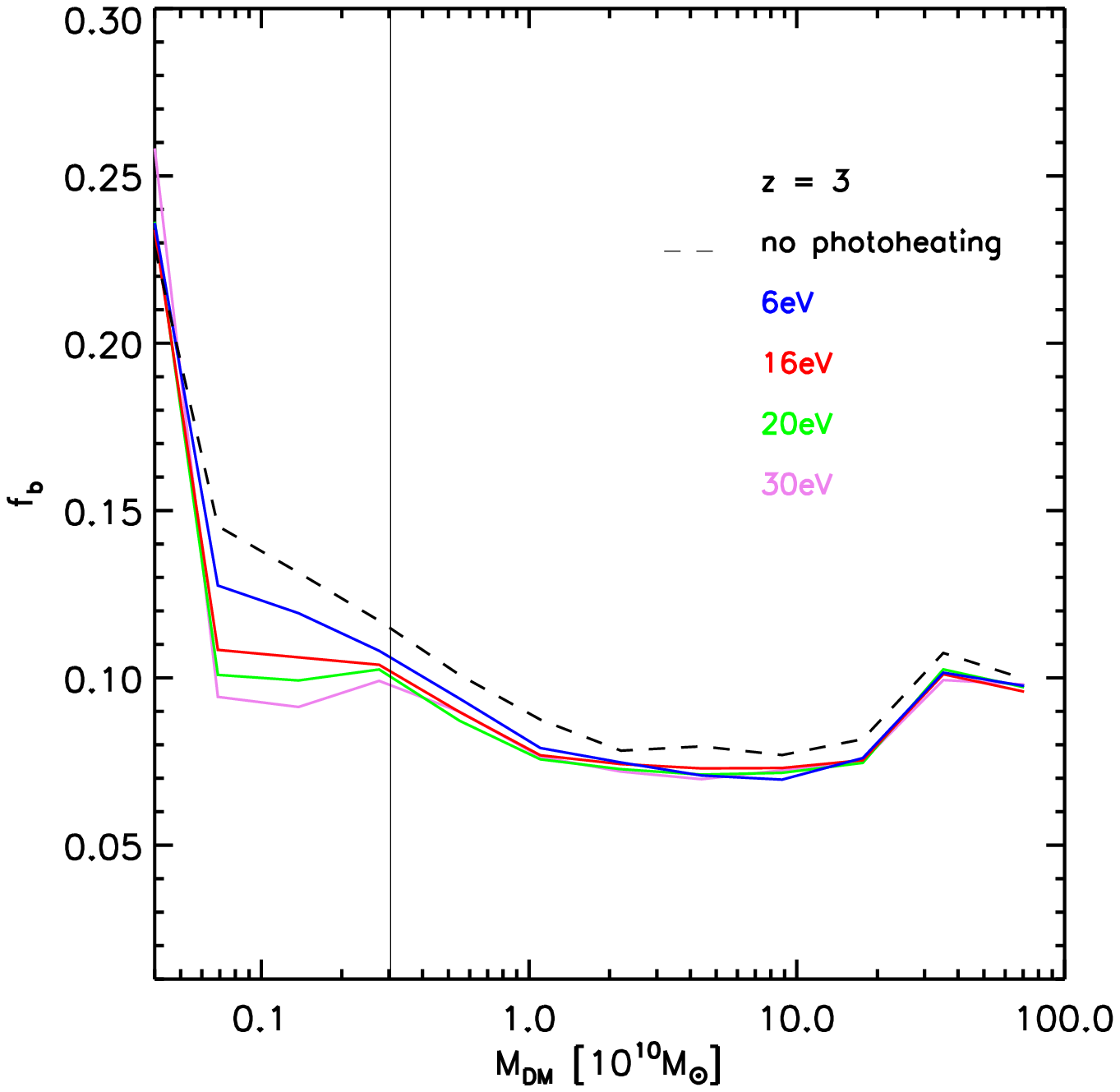}
\caption{Baryon fraction as a function of DM halo mass at $z=3$ in the
  lower resolution simulations with efficiency $\eta = 0.1$, and for
  different averaged photon excess energy. The dashed line shows the
  baryon fraction when no photo-heating has taken place. The black
  vertical line corresponds to a mass of 100 DM particles. Photo-heating
  does not affect the baryon fraction as strongly as in the high
  resolution simulation. We also observe that differences in the excess
  photon energy do not have a large effect on the baryon fraction.
\label{fig:f_b_esc010}} \efig

In Figure~\ref{fig:temperature}, we explore the temperature evolution of
the gas at different characteristic densities, corresponding to
under-dense gas by a factor of 10, gas at the mean density, and over-dense
gas by a factor of 10 relative to the mean. We compare our default
simulation with radiative transfer and photo-heating to the fiducial
simulation where no such heating is included at all. Clearly, the effect
of photo-heating is most prominent in the lowest density gas. This gas
is only weakly heated by structure formation shock waves when
photo-heating is not included. In contrast, when reionisation is
accounted for, the temperature of this gas reaches a high value of $\sim
10^4\,{\rm K}$ at the end of the epoch of cosmic reionisation, and even
before that, the mean temperature of this gas is raised considerably as
a result of the patchy and temporally extended reionisation transition
in our radiative transfer simulations. Interestingly, after reionisation
is complete, the mean temperature of the under-dense gas starts to slowly
decline again, while already for the mean density gas structure
formation shocks can provide for a slow further increase of the
temperature.

We also analysed the median temperature of the gas as a function
of over-density. As shown in Figure~\ref{fig:temp_inv}, after
reionisation has been completed, the low density gas ends up with a
higher median temperature than the higher density gas (except for the
gas in the star-forming phase). This points towards an `inverted
equation of state', as observed by \citet{Bolton2008},
\citet{Trac2008} and \citet{Furlanetto2009}. At later times, the
equation of state reverts again to a normal positive slope, when the
low density gas cools down due to the adiabatic expansion of the
Universe.

Finally, in Figure~\ref{fig:GSvsDM}, Figure~\ref{fig:f_b_256} and
Figure~\ref{fig:f_b_esc010} we explore the impact of the ionising
radiation field on the gas and stellar mass content of individual dark
matter halos.  To this end we run a group finder on our simulations and
simply determine the average gas mass, stellar mass and baryon fraction
of halos as a function of their dark matter mass.  We compare the $z=3$
results of our higher resolution radiative transfer simulation with the
simulation where photo-heating is completely ignored. Interestingly, we
find a reduction of the gas and stellar mass for all halo masses when
radiative transfer is included. The effect is quite weak for large halos
but becomes progressively larger for small halos. At dark matter halo
masses of $M_{\rm DM}= 10^9\,{\rm M}_\odot$ the suppression in baryonic
content is approximately 60\%, while at $M_{\rm DM}= 10^{12}\,{\rm
  M}_\odot$ it drops to only a few percent. This shows clearly the
important impact of the ionising radiation field on small dwarf
galaxies, in particular. While an externally imposed UV background can
perhaps account for the mean effect of this radiative feedback process
\citep{Hoeft2006,Okamoto2008}, only a spatially resolved treatment of
radiative transfer can account for effects of proximity that may well
play an important role in shaping, e.g., the satellite luminosity
function \citep{Munoz2009,Busha2010,Iliev2010}.

\section{Discussion and conclusions} \label{sec:conclusion}

We have presented the first application of our new implementation
\citep{Petkova2009} of radiation hydrodynamics in the cosmological
simulation code {\small GADGET}. We focused on the problem of cosmic
reionisation, aiming in particular at a first test on whether the
default star formation model in the code combined with our radiative
transfer modelling can yield a plausible reionisation history of the
Universe and a reasonable thermal state of the intermediate redshift
intergalactic medium. For simplicity, we have here only studied
star-forming galaxies as ionising sources, and restricted the analysis
to hydrogen reionisation alone. Based on the encouraging results
collected here, it is clearly worthwhile to extend the model further in
future work.

Since the level of internal absorption in the interstellar medium is
uncertain and cannot be resolved by our simulations, we have examined
models with different effective source efficiencies $\eta$. Likewise, as
we have not included a detailed spectral treatment and the time
evolution of non-equilibrium in the chemistry may be inaccurate, we have
parametrised the heat input per ionisation event in terms of a
parameter $\tilde \epsilon$.  

We find that our simulated universes can get reionised by
star formation in ordinary galaxies alone, with the epoch of
reionisation ending between redshifts $z=8$ to $z=5$, depending on the
assumed escape efficiency. The final phase transition is always quite
rapid in this our setup, but sizable fractions of the volume begin to be reionised much
earlier.  The heating efficiency has only a weak influence on the
reionisation history, but a stronger one on the cosmic star formation
rate density. In fact, we have shown that photo-heating plays an
important role in the evolution of the baryonic gas. As a result of the
associated heating, it changes baryonic structure formation by slowing
down the collapse of gas in dark matter halos, thereby delaying star
formation. This effect is strongest for the lowest mass halos, where the DM
potential well is not deep enough to easily overcome the thermal
pressure from the effects of photo-ionisation.

Our simple models of a self-consistent treatment of galaxy
formation and radiative transfer are not only able to produce a
plausible history of reionisation, but they also manage to approximately
match the basic statistics of the Lyman-$\alpha$ forest, at least for an
appropriate choice of the parameters $\eta$ and $\tilde \epsilon$. This
suggests that the low-redshift IGM data can be a powerful additional
constraint on future reionisation modelling in structure formation
simulations.

Despite these encouraging results it is also clear that our simulation
results are likely still affected by numerical resolution effects,
because the resolution in the lowest mass halos is still too coarse to
yield fully converged results. Ideally, we would like to resolve the
full range of star-forming halos with enough particles to achieve fully
converged results. While this is unlikely to qualitatively change any of
the results presented here, future precision work will require such
calculations. Another important caveat that will require further study
are uncertainties due to the radiative transfer approximation
itself. This is probably best addressed by comparing the results with a
completely different approach to radiative transfer. We are presently
finalising such a scheme in the moving-mesh code {\small AREPO}.  It
avoids the use of the diffusion approximation and can cast sharp
shadows, hence a direct one-to-one comparison with the optically-thin
variable Eddington tensor approach applied here will be particularly
interesting.

\section*{Acknowledgments}
The authors would like to thank Benedetta Ciardi for the helpful
discussion and comments. This research was supported by the DFG
cluster of excellence ``Origin and Structure of the Universe''
(www.universe-cluster.de).

\bibliographystyle{mn2e}
\bibliography{paper}

\label{lastpage}

\bsp

\end{document}